\documentclass[arps,twocolumn,showpacs,nofootinbib]{revtex4}

\usepackage{bm}
\usepackage{graphicx}
\usepackage{subfigure}
\input epsf

\newcommand\lsim{\mathrel{\rlap{\lower4pt\hbox{\hskip1pt$\sim$}}
        \raise1pt\hbox{$<$}}}
\newcommand\gsim{\mathrel{\rlap{\lower4pt\hbox{\hskip1pt$\sim$}}
        \raise1pt\hbox{$>$}}}

\def\tG{\tilde G {}}
\def\tT{\tilde T {}}
\def\tt{\tilde t {}}
\def\tnabla{\mbox{$\tilde \nabla$}}

\def\tPhi{\tilde \Phi {}}
\def\tPsi{\tilde \Psi {}}

\newcommand\ee{\end{equation}}
\newcommand\be{\begin{equation}}
\newcommand\eea{\end{eqnarray}}
\newcommand\bea{\begin{eqnarray}}
\newcommand{\sfrac}[2]{{\textstyle\frac{#1}{#2}}}
\newcommand\di{\partial}
\newcommand\mpl{M_{\rm Pl}}
\usepackage{amssymb}
\usepackage{amsmath}

\begin{document}
\title{Equivalence Principle Implications of Modified Gravity Models}

\author{Lam Hui}  
\email{lhui@astro.columbia.edu}
\affiliation{
Institute for Strings, Cosmology and Astroparticle Physics (ISCAP),\\
Department of Physics, Columbia University, New York, NY 10027, U.S.A.}

\author{Alberto Nicolis}
\email{nicolis@phys.columbia.edu}

\affiliation{
Institute for Strings, Cosmology and Astroparticle Physics (ISCAP),\\
Department of Physics, Columbia University, New York, NY 10027, U.S.A.} 

\author{Christopher W.~Stubbs}
\email{stubbs@physics.harvard.edu}
\affiliation{Department of Physics, Harvard University, Cambridge, MA 02138, U.S.A.}


\date{\today}

\begin{abstract}
Theories that attempt to explain the observed cosmic acceleration by modifying
general relativity all introduce a new scalar degree of freedom that is active
on large scales, but is screened on small scales to match experiments.
We demonstrate that if such screening occurs via the chameleon mechanism,
such as in $f(R)$ theory, it is possible to have order {\it unity} violation
of the equivalence principle, despite the absence of explicit violation
in the microscopic action. Namely, extended objects such as galaxies or
constituents thereof do not all fall at the same rate. 
The chameleon mechanism can screen
the scalar charge for large objects but not for small ones
(large/small is defined by the depth of the gravitational potential, and
is controlled by the scalar coupling).
This leads to order one fluctuations in the ratio of the inertial mass to
gravitational mass. We provide derivations in both Einstein and Jordan frames.
In Jordan frame, it is no longer true that all objects move on geodesics; only
unscreened ones, such as test particles, do.
In contrast, if the scalar screening occurs via strong coupling,
such as in the DGP braneworld model, equivalence principle violation occurs
at a much reduced level. We propose several observational tests of the
chameleon mechanism: 1. small galaxies should accelerate faster than large galaxies, 
even in environments where dynamical friction is negligible; 2. voids defined
by small galaxies would appear larger compared to standard expectations;
3. stars and diffuse gas in small galaxies should have different velocities, 
even if they are on the same orbits; 4. lensing and dynamical mass estimates
should agree for large galaxies but disagree for small ones.
We discuss possible pitfalls in some of these tests. The cleanest
is the third one where mass estimate from HI rotational velocity could exceed that
from stars by $30 \%$ or more. To avoid blanket screening of all objects, 
the most promising place to look is in voids.
\end{abstract}



\maketitle

\section{Introduction}
\label{intro}

The surprising finding of cosmic acceleration about ten years ago
has motivated a number of attempts to modify general relativity (GR) on large scales.
They fall roughly into two classes. One involves adding curvature invariants
to the Einstein-Hilbert action, the most popular example of which is $f(R)$ theory 
\cite{Carroll:2003wy,Nojiri2003,Carroll:2004de,Sotiriou2008}.
The other involves giving the graviton a mass or a resonance width, the most
well known example of which is the Dvali-Gabadadze-Porrati (DGP) braneworld model \cite{DGP,Deffayet}.
The first class of theories is equivalent to the classic scalar-tensor theories \cite{Chiba1,Chiba2}.
The second class of theories is more subtle but also reduces to scalar-tensor theories
on sub-Hubble scales \cite{LPR,NR}.
In the context of DGP, the extra scalar can be thought of as
a brane-bending mode. In fact, it appears the extra dimension is not even necessary---four dimensional 
ghost-free generalizations of the DGP model exist where it is
the nonlinear interaction of the scalar that causes self-acceleration \cite{NRT}.

All examples to date, therefore, introduce a light scalar (or even multiple light fields)
to modify gravity \footnote{This is not surprising given a famous theorem due to Weinberg \cite{WeinbergGR}
which states that a Lorentz invariant theory of a massless spin two field must be equivalent to GR
in the low energy limit. Low energy modifications of gravity therefore necessarily involve
introducing new degrees of freedom, such as a scalar. Giving the graviton a small mass or resonance width
turns out to introduce a scalar {\it \`a la} Stueckelberg \cite{AGS}. 
Theories that invoke a Lorentz violating massive graviton
also involve a scalar, namely a ghost condensate
\cite{ghostcondensate1,ghostcondensate2}.
Recent proposals for degravitation are yet another example of scalar-tensor theories in
the appropriate limits \cite{degrav,BHHK}.
}.
While such a scalar is welcome phenomenologically on cosmological scales,
it must be suppressed or screened on small scales to satisfy stringent constraints from
solar system and terrestrial experiments.
The two classes of theories have quite distinctive screening mechanisms.
Curvature-invariant theories such as $f(R)$ screen the scalar via the so-called
chameleon mechanism, essentially by giving the scalar a mass that depends on
the local density \cite{mota,chameleon1,chameleon2,cembranos2005,Hu:2007nk,
Tsujikawa2007,Cognola2007}. 
Theories such as DGP screen the scalar {\it \`a la} Vainshtein \cite{Vainshtein}, 
namely strong coupling effects suppress the scalar on small scales \cite{LPR,NR}.

Our goal in this paper is to study how extended objects move in the presence
of these two screening mechanisms. We will keep our discussion as general as possible
even as we use $f(R)$ and DGP as illustrative examples. Indeed, neither
theory is completely satisfactory: acceptable $f(R)$ models accelerate
the universe mostly by a cosmological constant rather than by a genuine
modified gravity effect \cite{Brax2008}; also from an effective field theory standpoint they are not favored in any sense over more general scalar-tensor theories; DGP and generalizations thereof \cite{NRT} probably  lack a relativistic ultra-violet completion \cite{AADNR}.
However, it is likely the screening mechanism, chameleon or strong coupling,
will remain relevant in future improvements of these theories.

The problem of motion of extended objects has a venerable history.
Newton famously showed in {\it Principia} that the Earth's motion around the Sun
can be computed by considering the Earth as a point mass.
It is worth emphasizing that such a result is by no means guaranteed
even in Newtonian gravity. This result holds only if there is a separation
of scales, that the dimension of the Earth is small compared with
the scale on which the Sun's gravitational field varies, in other words that
tides from the Sun can be ignored. More concretely, the Earth effectively sees a linear
gradient field from the Sun; second derivatives of the Sun's gravitational potential
can be safely neglected when computing the Earth's motion.

We will work within the same zero-tide approximation in this paper.
It has long been known that within GR, under the same approximation,
extended objects move just like infinitesimal test particles. In other words,
neglecting tidal effects all objects move on geodesics, or equivalently, their inertial mass and gravitational
mass are exactly equal. This has been
referred to as the effacement principle, that the internal structure of an object,
no matter how complicated, has no bearing on its overall motion \cite{Damour}.

It is also well known that the same is not true in scalar-tensor theories.
In such theories, despite the universal coupling of the scalar to all elementary matter fields,
objects with different amounts of gravitational binding energy will move differently.
Effectively, they have different ratios of the inertial mass to gravitational mass, 
known as the Nordtvedt effect \cite{Nordtvedt1968}.\footnote{By comparing
the free fall of the Earth and the moon towards the sun, lunar laser ranging data
already place limits on such strong equivalence principle violation at the level of $10^{-4}$ 
\cite{williams}, with a factor of ten improvement expected in the near future
\cite{battat}. In this paper, we are interested in theories that
respect these solar system bounds while exhibiting $O(1)$ deviation from GR
in other environments.}
This apparent violation of the equivalence principle
\footnote{In this paper, equivalence principle violation means no more and no less than this:
that different objects can fall at different rates, or equivalently, they have
different ratios of inertial to gravitational mass. See \cite{Sotiriou2007}
for a discussion of different formulations of the equivalence principle.}
is generally small because most
objects have only a small fraction of their total mass coming from their gravitational binding
energy. 
In other words, such violation is suppressed by $\Phi$ where $\Phi$ is the gravitational potential 
depth of the object. In the parlance of post-Newtonian expansion, such violation is an order
$1/c^2$ effect ($c$ is the speed of light).
Order one violation of the equivalence principle 
is however observable in the most strongly bound objects i.e.
black holes, which do not couple to the scalar at all---they have no scalar hair.
This means black holes and stars fall at
appreciably different rates in scalar-tensor theories \cite{Will1989,Will2006}.

In this paper, we wish to point out that there could be order one 
violation of the equivalence principle, even for objects that are not strongly bound,
such as galaxies and most of their constituents
($|\Phi| \ll 1$). In the parlance of post-Newtonian expansion, this is an order $1$ 
effect (i.e. this is not even {\it post-}Newtonian!).
That this is possible is thanks to the new twist introduced by recent variants of
scalar-tensor theories, namely their screening mechanisms.
We will show that the chameleon mechanism leads to order one violation of
the equivalence principle even for small $|\Phi|$ objects, while the Vainshtein mechanism
does not. These screening mechanisms were absent in traditional treatments of
scalar-tensor theories because the latter typically did not have a non-trivial scalar potential 
(needed for chameleon) or higher derivative interactions (needed for Vainshtein) i.e. they were of the 
Jordan-Fierz-Brans-Dicke variety \cite{Fierz,Jordan,BD1961}.

How such equivalence principle violation comes about is easiest to see in
Einstein frame, where the scalar mediates an actual fifth force.
Under the chameleon mechanism, screened objects effectively have a much smaller 
scalar charge compared to unscreened objects of the same mass. Hence, they respond differently to
the fifth force from other objects in the same environment; they then move differently.
That such a gross violation of the equivalence principle is possible 
might come as a surprise in Jordan frame.
One is accustomed to thinking that all objects move on geodesics in Jordan frame.
One of our main goals is to show that this is in fact not true.
Whether two different objects fall at the same rate under gravity
is of course a frame-independent issue; nonetheless, some readers
might find a Jordan frame derivation illuminating.
Here, we follow the elegant methodology developed by 
Einstein, Infeld \& Hoffmann \cite{EIH} and
Damour  \cite{Damour} and apply it to situations
where the screening mechanisms operate. What we hope to accomplish in part is putting
the arguments of Khoury and Weltman \cite{chameleon1,chameleon2} on a firmer footing, 
who first pointed out
the existence of the chameleon mechanism, and used Einstein frame
arguments to propose equivalence principle tests using
artificial satellites. 
We also wish to establish that if screening occurs via the Vainshtein mechanism,
there is no analogous order one violation of equivalence principle.

It is worth emphasizing the different physical origins for
equivalence principle violations of the post-Newtonian $O(1/c^2)$ 
type and
the Newtonian $O(1)$ type.
In any scalar-tensor theory, the scalar is always sourced by 
the trace of the matter energy-momentum tensor. In other words, at the level
of the Einstein frame action,
the coupling between the scalar $\varphi$ and matter 
is of the form $\varphi T^m$, where $T^m$ is the trace of the matter energy-momentum.
Fundamentally, equivalence principle violations occur 
whenever different objects couple to $\varphi$ differently. 
In a theory like Brans-Dicke, this happens because different objects of the same 
mass have different $T^m$. The degree to which their $T^m$
differs is of order $1/c^2$ (i.e. $\sim \Phi$ the gravitational potential), and 
so equivalence principle violation of
the Brans-Dicke type is typically small, except for relativistic objects whose
$T^m$ is highly suppressed i.e. most of their mass comes from gravitational binding energy.
An extreme example is the black hole whose mass is completely gravitational in origin.
In theories where the chameleon mechanism operates, there is an additional
and larger source of equivalence principle violation, and that is through variations in
the effective scalar charge that a macroscopic object carries. The scalar self-interaction
in a chameleon theory makes the scalar charge sensitive to the internal structure of the
object; the scalar charge is not
simply given by a volume integral of $T^m$ as in Brans-Dicke, but is given by a screened
version of the same. As we will see, chameleon screening could introduce $O(1)$ fluctuations in
the scalar charge to mass ratio, even for non-relativistic objects.
What is perhaps surprising is that
this happens despite an explicitly universal scalar coupling at the level
of the microscopic action. One can view this as a classical renormalization of
the scalar charge.
We will also see that in theories where the Vainshtein mechanism operates,
there is no such $O(1)$ charge renormalization. Therefore these theories only have
equivalence principle violation of the Brans-Dicke, or post-Newtonian, type.

Equivalence principle violations of both types discussed above, Newtonian or post-Newtonian,
should be clearly distinguished from yet another type of violation, which is
most commonly discussed in the cosmology literature --- where there is explicit breaking of
equivalence principle at the level of the microscopic action, e.g. elementary baryons and dark
matter particles are coupled to the scalar with different strengths 
\cite{Frieman,Stubbs,farrarpeebles,Gubser,Bean,keselman}. In this paper, we are interested in
theories where this breaking is not put in 'by hand' i.e. it does not exist at the level
of the microscopic action; rather, the equivalence principle violation comes about
for macroscopic objects by way of differences in response to the screening mechanism.

Further experimental tests come to mind after our investigation of the problem of
motion under the two screening mechanisms. We propose several observational tests
using either the bulk motion of galaxies or the internal motion of objects within galaxies.
The chameleon mechanism will predict gross violations of the equivalence principle for
suitably chosen galaxies while the Vainshtein mechanism will not.
Care must be taken in carrying out some of these tests, however, to avoid possible
Yukawa suppression and/or confusion with astrophysical effects. We point out that the
most robust test probably comes from the measurement of internal stellar and HI velocities 
of small galaxies in voids. Systematic differences between the two, or lack thereof, will
put interesting constraints or rule out some of the current modified gravity theories.

{\it A guide for the readers.} Readers who are interested primarily in observational
tests can skip directly to \S \ref{tests} where relevant results from the more theoretical earlier 
sections are summarized. \S \ref{GR} serves as a warm-up exercise wherein we derive the equality
of inertial mass and gravitational mass in GR for extended objects. 
\S \ref{QFT} demystifies this surprising effacement property of GR by viewing GR through
the prism of a field theory of massless spin two particles. This section is somewhat
of a digression and can be skipped by readers not interested in the field theoretic
perspective.
\S \ref{chameleon} is where we deduce the motion of extended objects under the chameleon mechanism
in Einstein frame. The Jordan frame derivation is given in the Appendix.
In \S \ref{strongcoupling} we do the same for the Vainshtein
or strong-coupling mechanism. Observational tests are discussed in \S \ref{tests}, where we also 
summarize our results.

A number of technical details are relegated to the Appendices.
Appendix \ref{app:multipoles} discusses how the requisite surface integral in
the Einstein-Infeld-Hoffmann approach can be done very close to the object.
Appendix \ref{app:fR} maps our parameterization of general scalar-tensor theories to
the special cases of Brans-Dicke and $f(R)$. Appendix \ref{app:jordan} provides the
Jordan frame derivation. Appendix \ref{forceVainshtein} derives in detail the equation of motion under
the Vainshtein mechanism. Appendix \ref{app:FRW} supplies useful results for an expanding universe.

A brief word on our notation: the speed of light $c$ is set to unity unless otherwise stated;
$8\pi G$ and $1/\mpl^2$ are used interchangeably; the scalar field $\varphi$ (or $\pi$) is
chosen to be dimensionless rather than have the dimension of mass.

\section{The Problem of Motion in General Relativity --- a Warm Up Exercise}
\label{GR}

We start with a review of an approach to the problem of motion introduced by
Einstein, Infeld \& Hoffmann \cite{EIH} and developed by Damour \cite{Damour},
using GR as a warm up exercise.
The idea is to work out the motion of an extended object
using energy momentum conservation, rather than to assume geodesic motion from the beginning.
It helps clarify the notion that the motion of extended objects is 
something that should be derived rather
than assumed---depending on one's theory of gravity, the extended objects might or might 
not move like infinitesimal test particles.

The Einstein equation ${G_\mu}^\nu = 8\pi G \, {T^m}_\mu {}^\nu$ can be rewritten as:
\begin{eqnarray}
\label{G1eqt}
{G^{(1)}}_\mu {}^\nu = 8 \pi G \, t_\mu {}^\nu
\end{eqnarray}
where ${G^{(1)}}_\mu {}^\nu $ is the part of the Einstein tensor ${G_\mu}^\nu$ that
is first order in metric perturbations, and
$t_\mu {}^\nu$ is the pseudo energy-momentum  tensor which is related
to the energy-momentum tensor ${T^m}_\mu {}^\nu$ by
\begin{eqnarray}
\label{tmunu}
t_\mu {}^\nu = {T^m}_\mu {}^\nu - {1\over 8\pi G} {G^{(2)}}_\mu {}^\nu 
\end{eqnarray}
where ${G^{(2)}}_\mu {}^\nu $ is the higher order part of ${G_\mu}^\nu$, including
but not necessarily limited to second order terms.
The superscript $m$ on ${T^m}_\mu {}^\nu$ is there to remind us this
is the matter energy-momentum, in anticipation of other kinds of energy-momentum
we will encounter later.
Note that the identity $\partial_\nu {{G^{(1)}}_\mu {}^\nu } = 0$ implies
$t_\mu {}^\nu$ is conserved in the ordinary (as opposed to covariant) sense.

Let us draw a sphere of radius $r$ enclosing the extended object whose motion we are
interested in. The linear momentum in the $i$-th direction is given by
\begin{eqnarray}
\label{Pidef}
P_i = \int d^3 x \, t_i {}^0
\end{eqnarray}
where the volume integral is over the interior of the sphere.
We assume that $t_i {}^0$ is dominated by the extended object such that
$P_i$ is an accurate measure of the linear momentum of the object itself, even
though we are integrating over a sphere that is larger than the object.

The gravitational force on the extended object is
\begin{eqnarray}
\label{force}
\dot P_i = \int d^3 x \, \partial_0 {t_i {}^0} = - \int d^3 x \, \partial_j {t_i {}^j}
= - \oint dS_j {t_i {}^j}
\end{eqnarray}
where in the last equality we have used the Gauss law to convert a volume
integral to a surface integral i.e. $d S_j = dA \, \hat x_j$, where
$dA$ is a surface area element and $\hat {x}$ is the unit outward normal.
As expected, the gravitational force is given by the integrated momentum flux through 
the surface.
This is the essence of the method introduced by Einstein, Infeld \& Hoffmann: 
the problem of motion
is reduced to performing a surface integral---i.e. there is no need to
worry about internal forces within the object of interest.

Assuming that ${{T^m}_i}^j$ at the surface of the sphere is
negligible, we need only consider the contribution
to $t_i {}^j$ from ${G^{(2)}}_i {}^j$ (Eq. (\ref{tmunu})).
And if the surface is located at a sufficiently large distance
away from the object of interest, we can assume the metric perturbations
are small and only second order contributions to
${G^{(2)}}_i {}^j$ need to be considered
(without assuming metric perturbations are small close to the object).
In the Newtonian gauge,
\begin{eqnarray}
ds^2 = - (1 + 2 \Phi) dt^2 + (1 - 2 \Psi) \delta_{ij} dx^i dx^j \; ,
\end{eqnarray}
we have:
\begin{eqnarray}
\label{G2ij}
&& {G^{(2)}}_i {}^j = 
- 2 \Phi (\delta_{ij} \nabla^2 \Phi - \partial_i \partial_j \Phi)
- 2 \Psi (\delta_{ij} \nabla^2 \Psi - \partial_i \partial_j \Psi) 
\nonumber \\
&& + \partial_i \Phi \partial_j \Phi - \delta_{ij} \partial_k \Phi \partial^k \Phi
+ 3 \partial_i \Psi \partial_j \Psi - 2 \delta_{ij} \partial_k \Psi \partial^k \Psi
\nonumber \\
&& - \partial_i \Phi \partial_j \Psi - \partial_i \Psi \partial_j \Phi 
+ 2\Psi {G^{(1)}}_i {}^j
\end{eqnarray}
ignoring time derivatives.
On the right hand side, we are being cavalier about
the placement of upper/lower indices---it does not matter as long
as we are keeping only second order terms.
The symbol $\nabla^2$ denotes $\partial_k \partial^k$.
The first order Einstein tensor ${G^{(1)}}_i {}^j$ is given by:
\begin{eqnarray}
\label{G1ij}
{G^{(1)}}_i {}^j = 
{\delta_{ij}} \nabla^2 (\Phi - \Psi) 
+ \partial_i \partial_j (\Psi - \Phi)
\end{eqnarray}
again, ignoring time derivatives.

The next step is to decompose $\Phi$ and $\Psi$ around the surface of the sphere
into two parts:
\begin{eqnarray}
\label{decompose}
\Phi = \Phi_0 + \Phi_1 (r) \quad , \quad 
\Psi = \Psi_0 + \Psi_1 (r)
\end{eqnarray}
where $\Phi_0$ and $\Psi_0$ represent the large scale fields due to the environment/background
while $\Phi_1$ and $\Psi_1$ are the fields due to the extended object itself.
This decomposition can be formally defined as follows:
$\Phi_1$ and $\Psi_1$ solve the Einstein equation with the extended object
as the one and only source; $\Phi_0$ and $\Psi_0$ are linear gradient fields
that can always be added to solutions of such an equation (because the equation
consists of second derivatives). We have in mind $\Phi_0$ and $\Psi_0$ are
generated by other sources in the environment, and they vary gently on the scale of the spherical
surface that encloses the object i.e. their second gradients can be ignored:
\begin{eqnarray}
\label{tide}
\Phi_0 ({\vec x}) \simeq \Phi_0 (0) + \partial_i \Phi_0 (0) x^i \nonumber \\
\Psi_0 ({\vec x}) \simeq \Psi_0 (0) + \partial_i \Psi_0 (0) x^i 
\end{eqnarray}
where $0$ denotes the center of the sphere. In other words, we assume
$\partial_i \Phi_0$ and $\partial_i \Psi_0$ hardly vary on the scale of the
sphere. On the other hand $\Phi_1$ and $\Psi_1$, the object generated fields,
have large variations within the sphere. At a sufficiently large $r$, we expect
both to be 
\be \label{monopole}
\Phi_1, \Psi_1 \simeq - GM/r
\ee
i.e. dominated by the monopole. 

Note that we have carefully chosen the radius $r$ of the enclosing sphere
to fulfill two different requirements: that $r$ is smaller than the scale
of variation of the background fields, and that $r$ is sufficiently large
to have the monopole dominate. The latter condition simplifies our calculation
but is not necessary in fact, as is shown in Appendix \ref{app:multipoles}. 
Also we have in mind a situation where the density within the
object of interest is much higher than its immediate environment---that is the definition of `object' after all---so that the total mass inside our sphere is dominated by the object's.

Plugging the decomposition (Eq. (\ref{decompose})) into 
Eq. (\ref{G2ij}) and performing the surface integral
as in Eq. (\ref{force}), we find
\footnote{We have {\it not}
assumed equality of $\Phi_0$ and $\Psi_0$, nor $\Phi_1$ and $\Psi_1$
in Eq. (\ref{G2integrate}). Keeping their relationship
general will be useful for the scalar-tensor case.
Eq. (\ref{G2integrate}) is derived
assuming only that $\Phi$ and $\Psi$ are subject to
an object-background split with a linear background. 
}
\begin{eqnarray}
\label{G2integrate}
&& \dot P_i = {1\over 8\pi G} \oint dS_j {G^{(2)}}_i {}^j \nonumber \\
&& \quad = {r^2 \over 2G} \, \partial_i \Phi_0 
\Bigl[ -{4\over 3} {\partial \Psi_1 \over \partial r}
-{2\over 3} {\partial \Phi_1 \over \partial r}\Bigr] \; .
\end{eqnarray}
This result has several remarkable features.
First, only cross-terms between the background fields
(subscript $0$) and the object fields (subscript $1$) survive
the surface integration.
A term such as $\oint dS_j \partial_i \Phi_0 \partial_j \Psi_0$
vanishes because by assumption, $\partial_i \Phi_0$ and
$\partial_j \Psi_0$ are both constant (or nearly so on
the scale of the sphere). 
A term such as $\oint dS_j \partial_i \Phi_1 \partial_j \Psi_1$
vanishes because both $\Phi_1$ and $\Psi_1$ are spherically symmetric.
Second, only terms proportional to $\partial_i \Phi_0$ remain.
There are no $\partial_i \Psi_0$ terms.
This should come as no surprise---by ignoring time derivatives, we
are in essence assuming the object is moving with non-relativistic speed
and so its motion should only be sensitive to the time-time part of
the background/environmental metric.

Finally, plugging the monopole $\Phi_1$ and $\Psi_1$ into
Eq. (\ref{G2integrate}) gives us:
\begin{eqnarray}
\label{MPhi0}
\dot P_i = - M \partial_i \Phi_0
\end{eqnarray}
which is the expected GR prediction in the Newtonian limit.

To complete this discussion, it is useful to note that
the mass of the object is given by
\begin{eqnarray}
M = - \int d^3 x \, {t_0}^0 \; .
\end{eqnarray}
Its time derivative $\dot M = \int d^3 x \, \partial_i {t_0}^i$
can once again be turned into a surface integral
$\dot M = \oint dS_i \,  {t_0}^i$. Assuming the energy flux through
the surface is small, we can approximate $M$ as constant.
The center of mass coordinate of the object is defined by
\begin{eqnarray}
X^i \equiv - \int d^3 x \, x^i \, {t_0}^0 / M \; .
\end{eqnarray}
Making the constant $M$ approximation, its time derivative is given by
\be
\label{Pdotz}
\dot X^i = \left[\int d^3 x \, \partial_j (x^i {t_0}^j) - {t_0}^i \right] / M
= - \int d^3 x \, {t_0}^i / M  \; .
\ee
This is precisely
$P_i/M$ defined in Eq. (\ref{Pidef}). Taking another time derivative, 
we can see that Eq. (\ref{MPhi0}) is equivalent to
\begin{eqnarray}
\label{FMA}
M \ddot X^i = - M \partial_i \Phi_0 \; .
\end{eqnarray}
As expected, the center of mass of an extended object moves on a geodesic. 
Its mass cancels out of the equation of motion i.e.~inertial mass equals gravitational mass.
That this holds independent of the internal structure of the extended object
is sometimes referred to as the effacement principle \cite{Damour}.

It is useful to summarize the approximations or assumptions we have made to obtain
this result, some of which are crucial, some of which not.

1. We have taken the non-relativistic/Newtonian limit i.e.~the object
has the equation of state of dust and is slowly moving, and
time derivatives of the metric are ignored.
These assumptions can be relaxed and the object would still move
on geodesics. For instance, photons surely move on geodesics in GR, but
we will not dwell on the proof here.

2. We assume the metric perturbations are small at the surface of the
sphere that encloses the object of interest. The derivation does not
assume however that metric perturbations are small close to the object.
For instance, Eq. (\ref{FMA}) remains valid for the motion of a black hole.
Note also that the assumption of a spherical surface is not crucial---the surface can take any shape, as discussed in Appendix \ref{app:multipoles}.

3. We assume there is a separation of scales, that external tides can be ignored
on the scale $r$ of the enclosing surface, and that the monopole contribution dominates at $r$. 
The last assumption can be easily relaxed. As shown in Appendix \ref{app:multipoles},
one could choose an $r$ that is close to the object, and the higher multipoles
would still not contribute to the relevant surface integral.
This is an important point because the enclosing surface is purely
a mathematical device---it should not matter what shape or size it takes, as long
as it is not so big that too much of the exterior energy-momentum is enclosed
or, equivalently, external tides become non-negligible.

Let us end this section with an interesting observation.
This derivation makes use of the metric that the extended object
sources (Eq. (\ref{monopole})). It is intriguing that what the object
{\it sources} also determines how the object {\it responds} to an 
external background field. In fact this is not surprising from the Lagrangian viewpoint, and is in a sense the modern generalization of Newton's third law: the same object-field interaction Lagrangian term yields both the source term in the field equations for gravity and the gravitational force term in the object's equation of motion.
This will be useful in understanding
how equivalence principle violation comes about in scalar-tensor theories.

\section{Field-theoretical considerations}\label{QFT}

The fact that when tidal effects are negligible, extended objects move along geodesics of the background metric, looks somewhat magical if derived as above but  is in fact forced upon us by Lorentz invariance.\footnote{
In this section, we digress a bit and wish to throw some light
on the effacement property of GR from a field-theoretical perspective.
Readers interested mainly in the problem of motion in scalar-tensor theories
can skip to the next section without loss in the continuity of the logic.}
In particular, it extends beyond the non-relativistic/weak-field regime discussed above. For example, a small black-hole falling in a very long-wavelength external gravitational field, will follow a geodesic of the latter.

This can be shown by imposing the requirement that the theory of massless spin-two particles (gravitons) interacting with matter be Lorentz invariant. In this section, we will thus abandon momentarily the geometric description of GR, and  treat gravity as a  theory of gravitons in Minkowski space. After all, from the modern quantum field theory
(QFT) viewpoint this approach is  
what legitimizes GR as a theory of gravity in the first place: GR is the only consistent  low-energy 
Lorentz-invariant theory of interacting massless spin-two particles.

So, let's describe gravitons in flat space via a tensor field $h_{\mu\nu}$.
We want to exploit the smallness of the object in question with respect to the typical variation-scale of the external gravitational field. What we can do then, is treat the object as a point-particle with trajectory $x^\mu(\tau)$, and couple it to the  graviton field in all possible ways allowed by symmetry. (Here the $x^\mu$'s are flat-space coordinates, and $\tau$ is flat-space proper-time: $d \tau^2 = - \eta_{\mu\nu} \, d x^\mu d x^\nu$.)
This yields an effective field theory valid for wavelengths much longer than the object's size. The important symmetry for us is just Lorentz-invariance. When applied to gravitons, this also demands invariance under gauge transformations
\begin{equation} \label{gauge}
h_{\mu\nu} \to h_{\mu\nu} + \partial_\nu \xi_\mu + \partial_\mu \xi_\nu \, \; ,
\end{equation}
see e.g.~ref.~\cite{WeinbergQFT1}. However at the order we will be working, this is not relevant. The low-energy effective action is
\bea 
S_{\rm eff} & = & - M \int \! d \tau + S_{\rm grav} [h_{\mu\nu}] \nonumber \\
& & - M \,  f \int \! d \tau \,  h_{\mu\nu} S^{\mu\nu} + \dots
\eea
The first line is just the sum of the free point-particle action and of the pure gravity action, which includes the graviton free action as well as all graviton self-interactions. In the second line, we explicitly display one interaction term between $h_{\mu\nu}$ and the point-particle. $S^{\mu\nu}$ is a Lorentz-tensor built out of the point-particle degrees of freedom: its trajectory $x^\mu(\tau)$, as well as possibly internal degrees of freedom such as spin, multipole moments, etc.; $f $ is a coupling constant. For convenience, we pulled out of $f$ an explicit factor of the particle's mass $M$. The dots stand for additional interactions, which involve either higher powers of $h_{\mu\nu}$, or derivatives thereof, or both. 

Now, Weinberg \cite{WeinbergEP} showed that for this effective theory to be Lorentz-invariant, in the low-energy graviton limit:
\begin{itemize}
\item[{\em i)}]
$S^{\mu\nu}$ must be totally independent of the internal degrees of freedom, and  given by
\be
S^{\mu\nu} = \dot x^\mu \dot x^\nu \; .
\ee

\item[{\em ii)}]
The coupling constant $f$ must be {\em universal}, that is, the same for all particles.
\end{itemize}
This is nothing but the equivalence ``principle'', which from this viewpoint is rather a theorem.
Indeed, since $f$ is a universal constant, we can absorb its value into a redefinition of $h_{\mu\nu}$. Let's therefore set it to $1/2$, so that the effective action becomes
\be \label{ppaction}
S_{\rm eff} =  - M \! \int \! d \tau \big( 1 + \sfrac12  h_{\mu\nu} \dot x^\mu \dot x^\nu \big) + S_{\rm grav} [h_{\mu\nu}]+ \dots
\ee
We recognize in parentheses the first order expansion of the point-particle action in GR:
\be
S^{\rm GR}_{\rm pp} = - M \! \int \! d \tau \sqrt{-g_{\mu\nu} \, \dot x^\mu \dot x^\nu }
\ee 
with metric $g_{\mu\nu} = \eta_{\mu\nu} + h_{\mu\nu}$.
In fact, one can show that because of the gauge-invariance (\ref{gauge}) (which is also a consequence of Lorentz-invariance), the non-linear interactions
between $h_{\mu\nu}$ and the point-particle (which make up part of the `dots' in eq.~(\ref{ppaction})), as well as the (low-energy) graviton self-interaction of $S_{\rm grav}$, must combine to yield the usual GR expressions. The remainder of  the `dots'---interactions involving higher derivatives of $h_{\mu\nu}$---encode non-minimal couplings of $h_{\mu\nu}$ with the particle's multipole moments \cite{goldberger}, which lead to tidal effects. These however are negligible for very long wavelength gravitational fields, their relative importance being suppressed by powers of $a/\lambda$, where $a$ is the object's size, and $\lambda$ the graviton's wavelength.

These are of course very well known notions. We reviewed them here, to contrast the very constrained situation of GR with a generic scalar-tensor theory. For GR, any object will source and respond to long-wavelength gravitational fields as if it were a point particle: the equivalence between inertial and gravitational masses is guaranteed by Lorentz invariance, and as a consequence it is totally independent of the object's internal structure, and of the forces that are keeping the object intact (including the gravitational one). It applies equally to macroscopic objects with non-relativistic structure (like e.g.~a very soft, very cold gas cloud) and to those with a very relativistic one (like e.g.~a black hole). Notice that the parameter $M$ appearing in eq.~(\ref{ppaction}) is the total physical mass of the object, which includes all classical and quantum contributions to it. Among the former we have, on top of the constituents' masses, the classical binding energy of a macroscopic system. In fact this is just the tree-level approximation of the quantum contributions, which are particularly relevant for microscopic systems where the bulk of the physical mass is quantum-mechanical in origin, like e.g.~for the proton. Weinberg's argument \cite{WeinbergEP} shows that upon quantum corrections, the inertial and gravitational masses of a quantum-mechanical system get renormalized by exactly the same multiplicative factor
\footnote{This is the gravitational analogue of a similar---but not quite as powerful---statement for electric charges: quantum corrections renormalize {\em all} charges by the same multiplicative factor, because charge renormalization can be thought of as wave-function renormalization for the photon field. See e.g.~ref.~\cite{WeinbergQFT1}.
}.

No analogous statement can be made for scalar couplings. Consider indeed a scalar-tensor theory for gravity.
Also assume that it is a `metric' theory, which means that there is a conformal frame (Jordan's) where matter-fields are all minimally coupled to the same metric. This is the closest we can get to implementing the equivalence principle. Now, upon demixing the scalar $\varphi$ from the graviton $h_{\mu\nu}$ (that is, in Einstein frame), this prescription corresponds to a direct coupling $\varphi \, {T^m}$
between the scalar and matter fields, where $T^m = {T^m}_\mu {}^\mu$ is the trace of
the matter energy-momentum tensor.
The problem with such a coupling is that $T^m$ is not associated with any conserved charges. 
That is, in the above point-particle approximation, extending GR to a scalar-tensor theory 
(in Einstein frame) amounts to adding an interaction term
\be
S_{\rm int} = Q \int \! d \tau \,  \varphi + \dots \; ,
\ee
as well as of course a Lagrangian for $\varphi$, its self-interactions, and its interactions with $h_{\mu\nu}$. The dots in $S_{\rm int}$ denote interactions involving derivatives of $\varphi$, which can be thought of as multipole interactions---negligible for a long-wavelength scalar field.
Now, the coupling $Q$ is the spatial integral of $T^m$ over the object's volume, and is {\em not} a conserved charge. So, for instance, if we consider a collapsing sphere of dust eventually forming a black-hole, its coupling to $h_{\mu\nu}$ (the mass $M$) is conserved throughout the collapse, whereas its coupling to $\varphi$ (the scalar charge $Q$) is not: as the dust particles become more and more relativistic, their $T^m$ approaches zero, even as
the total mass $M$ remains constant, dominated more and more by gravitational binding energy. 
In fact we already know that for a black-hole, which is the end product of our example, $Q$ {\em must} be zero, because of the no-hair theorem \cite{bekenstein}: a black-hole cannot sustain a scalar profile, which in our language means that in the point-particle approximation it cannot couple to a long-wavelength $\varphi$. So, in comparing how a black-hole and a non-relativistic gas cloud fall in a very long-wavelength scalar-tensor gravitational field, 
we expect order one
  violation of the equivalence principle: the black-hole will {\em not} follow a geodesic of the 
Jordan-frame metric.

In general, even if we start with a universal coupling $\varphi \, T^m$ at the microscopic level, the ratio $Q/M$ for extended objects will be object dependent, and sensitive to how relativistic the object's internal structure is: it will be one for very non-relativistic objects (with a suitable normalization of $\varphi$), and zero for black holes. 
This kind of variation in $Q/M$, which is relativistic in nature ($O(1/c^2)$ in a post-Newtonian
expansion), is available to all scalar-tensor theories.
As we will see, theories where the chameleon mechanism is at work possess an additional
and more dramatic source of $Q/M$ variation---it can be $O(1)$ even for non-relativistic objects.
This screening mechanism arises from a particular kind of scalar self-interaction,
and it results in a scalar charge $Q$ not just determined by an integral of $T^m$, but also by the precise non-linear dynamics of the scalar itself close to the object. Essentially, the scalar field profile sourced by the object at short distances contributes, via non-linearities, to the coupling of the object to a long-wavelength external scalar. (As we will see, for the Vainshtein mechanism this scalar charge renormalization does not occur.)

The lesson is this: from the low-energy point-particle action viewpoint, the charges $Q$ for different objects 
are free parameters, and are not forced to be related to the corresponding total inertial masses $M$ 
in any way. Therein lies the seed for equivalence principle violation.


\section{The Problem of Motion in Scalar-Tensor Theories --- Screening by the Chameleon Mechanism}
\label{chameleon}

The fact that different objects might fall at different rates in 
scalar-tensor theories is easiest to see in Einstein frame---the scalar
$\varphi$ mediates a fifth force and the equivalence principle is violated whenever
the scalar charge to mass ratio fluctuates between objects.
On the other hand, that this should be true in Jordan frame
might come as a bit of a surprise: after all, are not all objects supposed to move on 
geodesics in Jordan frame? We therefore present two different derivations, one in
each frame. The Einstein frame derivation is presented in \S \ref{einstein}.
The Jordan frame derivation is briefly summarized in \S \ref{jordan} with details
given in Appendix \ref{app:jordan}.
In addition, Appendix \ref{app:fR} describes how to map our description
of general scalar-tensor theories to the special cases of Brans-Dicke and $f(R)$.

The Einstein frame action is
\begin{eqnarray}
&& S = \mpl^2 \int d^4 x \sqrt{-\tilde g}
\left[\sfrac12 \tilde R - \sfrac12 \tilde \nabla_\mu \varphi
\tilde \nabla^\mu \varphi - V(\varphi) \right] \nonumber \\
&& \quad + \int d^4 x \, {\cal L}_m (\psi_m, \Omega^{-2}(\varphi) \tilde g_{\mu\nu})  \; .
\label{EF}
\end{eqnarray}
Here, $\tilde{\,}$ denotes quantities in Einstein frame,
$\tilde R$ is the Einstein frame Ricci scalar, $\tilde \nabla_\mu \varphi
\tilde \nabla^\mu \varphi$ is supposed to remind us that the contraction is
performed with the Einstein frame metric $\tilde g_{\mu\nu}$ (there is otherwise
no difference between $\tilde \nabla_\mu \varphi$ and $\nabla_\mu \varphi$: 
$\tilde \nabla_\mu \varphi = \nabla_\mu \varphi = \di_\mu \varphi$). 
The symbol $\varphi$ denotes some scalar that would contribute to a fifth force. 
Upon a field-redefinition, its kinetic term can always be written as in eq.~(\ref{EF}).
For notational convenience for what follows, we decide to factor an $\mpl^2$ out of the whole scalar-tensor gravitational action. With this normalization $\varphi$ is dimensionless.  We assume $\varphi$ has a potential $V(\varphi)$, which given our normalization has  mass-dimension two.
The symbol $\psi_m$ denotes some matter field whose precise nature is not
so important; the most important point is that it is minimally coupled not
to $\tilde g_{\mu\nu}$ but to $\Omega^{-2}(\varphi) \tilde g_{\mu\nu}$, which is
precisely the Jordan frame metric:
\begin{eqnarray}
\label{conformaltransf}
g_{\mu\nu} = \Omega^{-2} (\varphi) \tilde g_{\mu\nu} \; .
\end{eqnarray}
Performing the conformal transformation gives us the Jordan frame action:
\begin{eqnarray}
\label{actionJordan}
&& S = \mpl^2 \int d^4 x \sqrt{-g} \Big[ \sfrac12 \Omega^2 (\varphi) R
- \sfrac12 h(\varphi) \nabla_\mu \varphi \nabla^\mu \varphi \nonumber  \\
&&  \qquad \qquad -\Omega^4(\varphi) V \Big] \nonumber \\
&& \quad + \int d^4 x \, {\cal L}_m (\psi_m, g_{\mu\nu})
\end{eqnarray}
where
\begin{eqnarray}
\label{hdef}
h(\varphi) \equiv \Omega^2 \left[1 - \frac32 \left({\partial {\,\rm ln\,} \Omega^2
\over \partial \varphi}\right)^2 \right]  \; .
\end{eqnarray}
The matter energy-momentum in the two frames (defined as the derivative of the action w.r.t.~the corresponding metric) are related by:
\begin{eqnarray}
{T^m}_{\mu\nu} = \Omega^2(\varphi) {\tT^m}_{\mu\nu} \; .
\end{eqnarray}



In anticipation of the fact that we will be carrying out perturbative computations,
let us point out that
$\Omega^2 (\varphi)$ must be close to unity for the metric perturbations
to be small in both Einstein and Jordan frames. We can thus approximate it at linear order in $\varphi$,
\begin{eqnarray}
\label{alphaphi}
\Omega^2 (\varphi) \simeq 1 - 2 \alpha \varphi \; ,
\end{eqnarray}
where $\alpha$ is a constant, and $|\alpha \varphi| \ll 1$. We are {\it not} assuming, however, that the fractional scalar field
perturbation (i.e. $\delta\varphi/\varphi$) is in any way small.
In fact, nonlinear effects in $\delta\varphi/\varphi$ are going to be crucial
in enabling the chameleon mechanism.
Under the same approximation, a useful relation is
\begin{eqnarray}
\label{alphastar}
{\partial {\,\rm ln\,} \Omega^2 \over \partial \varphi} \simeq -2\alpha \; .
\end{eqnarray}
The reader might find it helpful to keep in mind a particularly common (if not particularly natural) form
of the conformal factor, $\Omega^2 = {\,\rm exp\,} [{-2 \alpha \varphi}]$ 
with constant $\alpha$, in which case Eq. (\ref{alphastar}) is exact
and Eq. (\ref{alphaphi}) is a Taylor expansion.

As elaborated in Appendix \ref{app:fR}, $\alpha = 1/\sqrt{6}$ in $f(R)$ models
and $\omega_{\rm BD} = (1 - 6 \alpha^2)/(4\alpha^2)$ in Brans-Dicke theory.

\subsection{Derivation in Einstein Frame}
\label{einstein}
The Einstein frame metric equation is
\begin{eqnarray}
\label{metriceqtEinstein}
\tG_\mu {}^\nu = 
8\pi G \left[ {\tT^m}_\mu {}^\nu +
{\tT^\varphi}_\mu {}^\nu
\right]
\end{eqnarray}
where
\begin{eqnarray}
\label{TmunuscalarEinstein}
{\tT^\varphi}_\mu {}^\nu
= \frac{1}{8\pi G} \left\{ \tnabla_\mu \varphi
\tnabla^\nu \varphi - {\delta_\mu}^\nu
\left[ \sfrac12 \tnabla_\alpha \varphi
\tnabla^\alpha \varphi + V \right] \right\} .
\end{eqnarray}
The scalar field equation is
\begin{eqnarray}
\label{scalareqtEinstein}
\tilde{\Box} \varphi = {\partial V \over
\partial \varphi} +
4\pi G {\partial {\,\rm ln \,} \Omega^2
\over \partial \varphi} {\tT^m}_\mu {}^\mu \; .
\end{eqnarray}
Following the derivation in the GR case (\S \ref{GR}), we
perform an object-background split, derive solutions to the object fields,
and compute an integral of momentum flux over an enclosing sphere
to find the net gravitational force on the extended object.

\subsubsection*{Object-background Split}

We work in the Newtonian gauge:
\begin{eqnarray}
ds^2 = - (1 + 2 \tilde \Phi) dt^2 + (1 - 2 \tilde \Psi) \delta_{ij} dx^i dx^j \; .
\end{eqnarray}
As before (Eq. (\ref{decompose})), we carry out a decomposition
of the fields at some radius $r$ away from the object:
\begin{eqnarray}
\label{decomposeEinstein}
\tilde \Phi = \tilde \Phi_0 + \tilde \Phi_1 (r) \, , \;
\tilde \Psi = \tilde \Psi_0 + \tilde \Psi_1 (r) \, , \;
\varphi = \varphi_0 + \varphi_1 (r)
\end{eqnarray}
where the subscript $0$ denotes the background/environmental fields and
the subscript $1$ denotes the object fields. The background fields are
assumed to be well approximated by linear gradients:
\begin{eqnarray}
\label{lineargradEinstein}
\tilde \Phi_0 ({\vec x}) \simeq \tilde \Phi_0 ({0}) + \di_i \tilde \Phi_0 ({0}) x^i
\nonumber \\
\tilde \Psi_0 ({\vec x}) \simeq \tilde \Psi_0 ({0}) + \di_i \tilde \Psi_0 ({0}) x^i
\nonumber \\
\varphi_0 ({\vec x}) \simeq \varphi_* + \di_i \varphi_0 (0) x^i
\end{eqnarray}
where $0$ denotes the origin centered at the object, and we use $\varphi_*$
to denote the background scalar field value there. The assumption here
is that these background fields sourced by the object's environment
vary on a scale much larger than $r$, where we will draw our enclosing sphere.

It is worth emphasizing that this object-background split needs only make sense
at the surface of the enclosing sphere (over which we will compute the integrated
momentum flux), but not necessarily at the location of the object itself.

\subsubsection*{Solutions for the Object Fields}

To work out solutions for the object fields, we first examine
the Einstein equation linearized in metric perturbations:
\begin{eqnarray}
\label{Einsteinframe1storder}
&& \nabla^2 \tilde \Psi = 4\pi G \, \tilde \rho 
- \sfrac14 \di_0 \varphi \,  \di^0 \varphi + \sfrac14 \di_i \varphi \,
\di^i \varphi + \sfrac12 V\nonumber \\
&& \partial_0 \partial_i \tilde \Psi = - \sfrac12 \di^0 \varphi  \, \di_i \varphi \nonumber \\
&& (\partial_i \partial_j - \sfrac13 \delta_{ij} \nabla^2)(\tilde \Psi - \tilde \Phi)
= \nonumber \\
&& \quad \quad \quad \di_i \varphi \, \di^j \varphi - \sfrac13 {\delta_i}^j
\di_k \varphi \,\di^k \varphi  \nonumber \\
&& \partial_0^2 \tilde \Psi + \sfrac13 \nabla^2 (\tilde \Phi - \tilde \Psi)
= \nonumber \\
&& \quad \quad \quad \sfrac16
\left( -\sfrac32 \di_0 \varphi \, \di^0 \varphi -\sfrac12 \di_k \varphi \, \di^k \varphi - 3 V \right)
\end{eqnarray}
where we have assumed the matter is non-relativistic, characterized only by
its energy density $\tilde\rho$.

We can ignore
all second order scalar field terms on the right hand side of these
equations. This is because as we will see shortly, $\varphi$ is already first order in $G$. Thus its contribution to Einstein's equations is on an equal footing with post-Newtonian corrections, which we are neglecting. In addition, we will ignore $V$ as well.
The rationale is that $G \tilde \rho \gg V$ inside the object of
interest, whereas outside the object, $V$ or any other sources
of energy-momentum are just small (in 
the sense that the total mass within the enclosing sphere is dominated by
the object). Essentially, we assume the metric in the vicinity
of the object is sourced mainly by matter rather than the scalar field:\footnote{
Under this assumption, the anisotropic stress vanishes. This is
emphatically an Einstein frame statement. The same will not be true
in Jordan frame.}
\begin{eqnarray}
\label{Einsteinframe1orderB}
\nabla^2 \tilde \Psi = 4\pi G\tilde\rho \quad , \quad
\nabla^2 (\tilde \Phi - \tilde \Psi) = 0 \; .
\end{eqnarray}
Under the same assumptions as in \S\ref{GR}, we get
\begin{eqnarray}
\label{monopoleEinstein}
\tilde \Phi_1 = \tilde \Psi_1 = - GM/r
\end{eqnarray}
where $M$ is the mass of the object.

As in the case of GR, we in essence assume there is a separation of
scales: that $r$ can be chosen small enough that second gradients
of the background tidal fields can be ignored, 
but large enough for the monopole to dominate.
The latter condition is helpful but not essential (see Appendix
\ref{app:multipoles}).

The next task is to compute the scalar field profile sourced by
the object. The scalar field equation (Eq. (\ref{scalareqtEinstein}))
can be written as
\begin{eqnarray}
\label{scalarEinstein}
\nabla^2 \varphi = {\partial V \over \partial\varphi}
+ \alpha \, 8\pi G \, \tilde \rho
\end{eqnarray}
where we have ignored time derivatives and corrections due to metric perturbations,
and we have used the approximation Eq. (\ref{alphastar}).
Here, it is crucial we do not assume $\delta\varphi/\varphi$ is small even though
$\varphi$ is. 
\begin{figure}[htb!]
\begin{center}
\includegraphics[width=7cm]{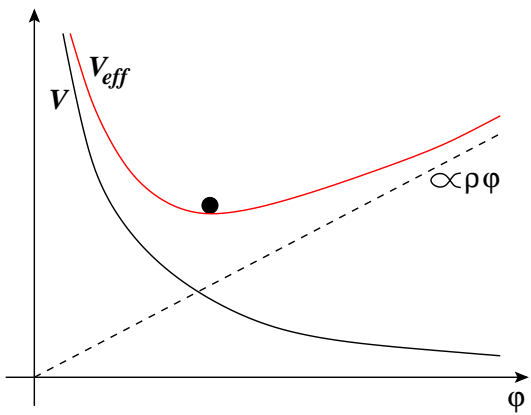}
\end{center}
\caption{\label{potential} A scalar potential for the chameleon mechanism. The effective potential felt by $\varphi$ in the presence of sources is the sum of the self-interaction potential
$V(\varphi)$ and the scalar-matter coupling $(\alpha \, 8 \pi G ) \tilde \rho \, \varphi$. 
This can give $\varphi$ a large mass at locations where the matter density $\tilde \rho$ is
large.}
\end{figure}
The important nonlinearity for the chameleon mechanism is in the potential term.
For instance, for the potential of fig.~\ref{potential}
the chameleon mechanism could operate,
meaning that for an object of the right size and density, 
$\varphi$ could be trapped at some small value 
inside much of the extended object---a value where
the two terms on the right hand side of Eq. (\ref{scalarEinstein})
roughly balance. This corresponds to $\varphi$
having a large mass, or a small Compton wavelength, inside the object
(see Fig. \ref{potential}).
The object will therefore
have an exterior scalar profile that is screened i.e.~only a fraction of the object's mass (the fraction
that happens to live within a shell approximately
a Compton wavelength thick at the object's boundary)
sources the exterior scalar profile.
In other words, we expect the object to produce a scalar field
that at large $r$ is dominated by a monopole of the form:
\begin{eqnarray}
\label{scalarmonopoleEinstein}
\varphi_1 (r) = - \epsilon  \, \alpha \, {2 G M \over r}
\end{eqnarray}
where $\epsilon$ quantifies the degree of screening.

If the scalar field remains light inside the object, we have
$\partial V/\partial \varphi \ll \alpha 8 \pi G \tilde \rho$ in which
case the exterior scalar profile is sourced by the full mass of the object,
i.e.~$\epsilon = 1$ in Eq. (\ref{scalarmonopoleEinstein}).
On the other hand, suppose the scalar field is massive 
and therefore short-ranged inside the object, the exterior scalar profile
is sourced only by the mass residing in a shell at the boundary of the object.
Suppose the object has a radius $r_c$ and the shell has thickness $\Delta r_c$.
The scalar field sits at some small equilibrium value (where the potential
and density terms on the right hand side of Eq. (\ref{scalarEinstein}) cancel)
at all radii interior to the object -- that is, except where the shell is.
At the shell, the scalar field starts to pull away from this small expectation value
because of the force exerted by the density term. In other words, at the shell,
Eq. (\ref{scalarEinstein}) can be approximated by
$\delta\varphi/(r_c \Delta r_c) \sim 
\alpha \, 8\pi G \, M/(4 \pi r_c^3/3)$,
where $\delta\varphi = \varphi_* - \varphi_c \simeq \varphi_*$ ($\varphi_c$ is
the equilibrium value where the the potential and density terms balance in the
deep interior of the object, and
$\varphi_*$ is approximately the scalar field value far away from the object).
We therefore have $3 \Delta r_c/ r_c \sim (\varphi_*/2\alpha) \cdot (GM/r_c)^{-1}$.
This is precisely the fraction by volume of the object that sources the 
exterior scalar field. This means the suppression factor $\epsilon$
in Eq. (\ref{scalarmonopoleEinstein}) should be
\begin{eqnarray}
\label{thinshell}
\epsilon \simeq {\varphi_* \over {2 \alpha}} \left[ {GM \over r_c} \right]^{-1}
\quad {\rm if} \quad 
{\varphi_* \over {2 \alpha}} \left[ {GM \over r_c} \right]^{-1} < 1
\end{eqnarray}
where $GM/r_c$ is the gravitational potential of the object.
The inequality is required for consistency: it had better be true that
$3\Delta r_c / r_c < 1$ for screening to occur. This is known
as the thin-shell condition. When this condition is violated, there is
no screening:
\begin{eqnarray}
\label{nothinshell}
\epsilon = 1 \quad {\,\rm if\,} \quad {\varphi_* \over {2 \alpha}} \left[ {GM \over r_c} \right]^{-1}
\gsim 1 \; ,
\end{eqnarray}
essentially meaning that $\alpha 8 \pi G \tilde \rho \gg {\partial V/\partial \varphi}$ {\it throughout} 
the object, in which case the scalar field $\varphi$ cannot even reach the equilibrium value $\varphi_c$
inside the object. 
It is interesting to note that the thin-shell condition does not
depend explicitly on the potential $V$.
The implicit requirement is that the potential is of such a form that
equilibrium (between density and potential) is possible.
A more detailed derivation of the thin-shell condition can be found in
\cite{chameleon2}.

Along the lines of \S\ref{QFT},
it is useful to introduce the idea of a scalar charge: by analogy with
$\tilde \Psi_1 = - GM/r$ for the gravitational potential, let us
write the scalar profile as $\varphi_1 = - 2 G Q/r$ (the factor of
$2$ is motivated by the appearance of $8\pi \tilde\rho$ rather than 
$4\pi\tilde\rho$ in Eq. (\ref{scalarEinstein})). By inspecting
Eq. (\ref{scalarmonopoleEinstein}), we can see that the dimensionless scalar charge $Q$ is
\begin{eqnarray}
\label{Qdef} 
Q \equiv \epsilon \, \alpha M \; .
\end{eqnarray}
Screened objects carry a much reduced scalar charge as viewed from outside.

Just as in the case of the metric fluctuations, one can always add
to $\varphi_1$ a linear gradient of the form
Eq. (\ref{lineargradEinstein}) to represent the field generated
by sources in the environment. 
To summarize, $\tilde \Phi$, $\tilde \Psi$ and $\varphi$
are given by Eqs. (\ref{decomposeEinstein}), (\ref{lineargradEinstein}),
(\ref{monopoleEinstein}) and (\ref{scalarmonopoleEinstein}).

\subsubsection*{Surface Integral of Momentum Flux}

We are ready to compute the gravitational force
following Eq. (\ref{force}):
\begin{eqnarray}
\dot P_i = - \oint dS_j \,  {\tt_i} {}^j
\end{eqnarray}
where the surface integral is again over some sphere enclosing the object of interest.
By analogy to Eqs. (\ref{G1eqt}) and (\ref{tmunu}), the pseudo energy-momentum tensor
is:
\begin{eqnarray}
\label{ttmunu}
\tt_\mu {}^\nu = {\tT^m}_\mu {}^\nu + {\tT^\varphi}_\mu {}^\nu 
- {1\over 8\pi G} {\tG^{(2)}}_\mu {}^\nu \; .
\end{eqnarray}
As before, ${\tT^m}_i{}^j$ outside the object is small,
and we have already computed the ${\tG^{(2)}}_i{}^j$
contribution (just put $\tilde {\,}$ on top of quantities in Eq. (\ref{G2integrate})).
What remains to be done is to compute the contribution from ${\tT^\varphi}_i{}^j$
(Eq. (\ref{TmunuscalarEinstein})).
The same tricks that work for the ${\tG^{(2)}}_i{}^j$ contribution
work here, and we obtain
\begin{eqnarray}
\label{Tscalarintegrate}
- \oint dS_j {\tT^\varphi}_i{}^j = - \frac{1}{2G} r^2 \, \partial_i \varphi_0 {\partial \varphi_1 \over \partial r}  \; .
\end{eqnarray}
Note that there should be a contribution from the potential of
the form: $\frac1G \oint dS_j V \sim \frac1G (4\pi r^3/3) \partial_i \varphi_0 
{\partial V / \partial \varphi |_{\varphi_*}}$, but this can be ignored by our assumption
that the scalar field $\varphi_1$ is sourced mainly by the object rather than its immediate environment.

Putting together all contributions (Eq. (\ref{Tscalarintegrate}) \& Einstein frame
analog of Eq. (\ref{G2integrate})), we therefore have
\begin{eqnarray}
\label{forceEinstein}
\dot P_i = {r^2 \over 2G} \bigg[ \partial_i \tilde \Phi_0 
\Bigl( -{4\over 3} {\partial \tilde \Psi_1 \over \partial r}
- {2\over 3} {\partial \tilde \Phi_1 \over \partial r}\Bigr)
-  \partial_i \varphi_0 {\partial \varphi_1 \over \partial r} \bigg] \; .
\end{eqnarray}

Putting our solutions for the various fields into the above expression, we obtain
\begin{eqnarray}
\label{Einsteinmain}
&& M \ddot X^i = - M \partial_i \tilde \Phi_0 - \epsilon \alpha M \partial_i \varphi_0
\nonumber \\ && \quad \quad = - M \left[ \partial_i \tilde \Phi_0 + {Q\over M} \, \partial_i \varphi_0 \right]
\end{eqnarray}
where have used Eq. (\ref{Qdef}), and we have equated $\dot P_i$ and $M \ddot X^i$, where
$X^i$ is the object center of mass coordinate (Eq. (\ref{Pdotz})).
This is the main result of this section. The fact that
screened and unscreened objects have different scalar charge to mass ratios $Q/M$
(or equivalently, different $\epsilon$) means they fall at different rates.
{\it This is the origin of apparent equivalence principle violation.}

This result can be further simplified if we make one more 
assumption, which does not necessarily hold in general.
Both $\tilde \Phi_0$ and $\varphi_0$ are sourced by
the environment. Suppose they are sourced in a similar way:
\begin{eqnarray}
\label{sourceenvironment}
\nabla^2 \tilde \Phi_0 = 4\pi G \tilde \rho \quad , \quad
\nabla^2 \varphi_0 = \alpha \, 8\pi G \, 
\tilde \rho
\end{eqnarray}
where $\tilde\rho$ is the environmental density. 
The latter equation follows  from Eq. (\ref{scalarEinstein})
provided that potential terms are small compared to density terms
on the scales of interest.
Eq. (\ref{sourceenvironment}) then implies
\begin{eqnarray}
\label{scalaePhi}
\varphi_0 = 2 \alpha \, \tilde \Phi_0 \; .
\end{eqnarray}
Putting this into Eq. (\ref{Einsteinmain}), we therefore obtain
\begin{eqnarray}
\label{Einsteinmain2}
&& M \ddot X^i = - M \partial_i \tilde \Phi_0 \left[ 1 + 2 \alpha {Q \over M} \right] \nonumber \\
&& \quad \quad \, \, = - M \partial_i \tilde \Phi_0
\big[1 + 2 \epsilon \alpha^2 \big] \; .
\end{eqnarray}
The fact that this result depends on the square of
$\alpha$ makes it clear that the sign of $\alpha$ has no
physical meaning---chameleon mechanism only requires its
sign be opposite to that of $\partial V/\partial\varphi$ (see 
Eq. (\ref{scalarEinstein})).

Suppose one has a scalar-tensor theory where
$2 \alpha^2$ is order unity, as is true
in $f(R)$ theories,
then screened objects ($\epsilon \sim 0$) and unscreened objects
($\epsilon = 1$) would move on very different trajectories i.e.~{\it order unity apparent violation of the equivalence principle}.
Note that infinitesimal test particles should go unscreened and therefore
have $\epsilon = 1$.

Let us end with a summary of further approximations we have made
in this section to augment those listed at the end of \S \ref{GR}:

1'. We have taken the non-relativistic or quasi-static 
limit i.e. time derivatives of the scalar field are ignored.

2'. We assume the scalar field does not contribute much to the r.h.s.~of Einstein's equations (\ref{Einsteinframe1storder}). 

3'. We assume there is a separation of scales, that second derivatives
of the background/environmental scalar field can be ignored on the scale $r$ of the enclosing
surface, 
and that the monopole contribution to the 
object's scalar profile dominates at $r$. As discussed in Appendix \ref{app:multipoles},
this latter assumption can be relaxed. 
The former  assumption implicitly 
requires the Compton wavelength of the exterior scalar field
to be large compared with $r$. The scalar field would be
Yukawa suppressed if this were not true. This has implications for
observational tests as we will discuss in \S \ref{tests}.

To this list, which mirrors the list in \S \ref{GR}, we should add:

4'. We assume the conformal factor $\Omega^2(\varphi)$ is close to unity, or
equivalently $|\alpha\varphi| \ll 1$ (see Eq. (\ref{alphaphi})). This assumption
is not strictly necessary in our Einstein frame calculation. 
It will become necessary when we try to compare the Jordan frame calculation
with it. One consequence of this assumption is that the Einstein frame and Jordan
frame masses of the object are roughly equal i.e.
\begin{eqnarray}
M = \int d^3 x \, \tilde \rho \simeq \int d^3 x \, \rho
\end{eqnarray}
since $\tilde \rho = \Omega^4 \rho$, recalling that
${\tT^m}_0 {}^0 = -\tilde \rho$ and ${T^m}_0 {}^0 = -\rho$.

In summary, the main results
of this section are encapsulated in Eqs. (\ref{forceEinstein}), 
(\ref{Einsteinmain}) and (\ref{Einsteinmain2}), which make 
different levels of approximations, starting
from the minimum. The key observation is that, despite the explicitly
universal coupling of the scalar to matter at the microscopic level 
(Eq. (\ref{EF})), scalar charges of macroscopic objects get classically renormalized
by nonlinear chameleon effects. 
This produces fluctuations in the scalar charge to mass ratio $Q/M$ for different objects,
leading to different rates of gravitational free fall.
Screened objects, which have $Q/M \sim 0$, accelerate less than unscreened objects.
As we will see, no such (non-relativistic) renormalization occurs for the Vainshtein mechanism.

\subsection{Derivation in Jordan Frame}
\label{jordan}

In principle, the simplest way to obtain the Jordan frame results is
to transform the end results of \S \ref{einstein}, Eqs. (\ref{forceEinstein}), 
(\ref{Einsteinmain}) and (\ref{Einsteinmain2}), into Jordan frame by a
conformal transformation $g_{\mu\nu} = \Omega^{-2} \tilde g_{\mu\nu}$.
After all, whether two different objects fall at different rates is a frame-independent issue.
This might not, however, satisfy readers who are accustomed to the Jordan frame
viewpoint: should not everything move on geodesics? 
In this section, we therefore briefly sketch a Jordan frame 
derivation. Details can be found in Appendix \ref{app:jordan}.

The Jordan frame metric in Newtonian gauge is
\begin{eqnarray}
ds^2 = - (1 + 2 \Phi) dt^2 + (1 - 2 \Psi) \delta_{ij} dx^i dx^j \; .
\end{eqnarray}
From Eq. (\ref{conformaltransf}) and recalling
$\Omega^2 \simeq 1 - 2\alpha\varphi$ (Eq. (\ref{alphaphi})),
the Jordan and Einstein frame metric perturbations are related by
\begin{eqnarray}
\label{smallconformal}
\Phi = \tilde \Phi + \alpha\varphi \quad , \quad \Psi = \tilde \Psi - \alpha\varphi \; .
\end{eqnarray}
Note how $\Phi \ne \Psi$ even if $\tilde \Phi = \tilde \Psi$.

The Jordan frame metric equation is
\begin{eqnarray}
\label{metriceqtJordan}
&& {G_\mu}^\nu = 8\pi G \,  \Omega^{-2}
\left[ {T^m}_\mu {}^\nu + {{T^\varphi}_\mu}^\nu \right] \nonumber \\
&& \quad + \Omega^{-2} \left[ \nabla_\mu \nabla^\nu \Omega^2 - {\delta_\mu}^\nu 
\Box \Omega^2 \right]
\end{eqnarray}
where
\begin{eqnarray}
{T^\varphi}_\mu{}^\nu & = & \frac{1}{8 \pi G} \Big\{
h \nabla_\mu \varphi \nabla^\nu \varphi \nonumber \\
& - &  {\delta_\mu}^\nu 
\Big[ \sfrac12 h \nabla_\alpha \varphi \nabla^\alpha \varphi
+ \Omega^4 V \Big] \Big\} \label{TmunuscalarJordan} \; .
\end{eqnarray}
The scalar field energy-momentum follows from (the minimally coupled
portion of) the Jordan frame action
Eq. (\ref{actionJordan}). Note that $\varphi$ does not have a canonical
kinetic term, hence the presence of $h(\varphi)$ (Eq. (\ref{hdef})).
In some theories, $h$ might even be zero, such as in $f(R)$.

The Einstein-Infeld-Hoffmann approach to the problem of motion is to
compute the gravitational force as a surface integral of the momentum flux
(Eq. (\ref{force})). The momentum flux can be obtained from the spatial
components of the pseudo-energy-momentum tensor, which is defined by Eq. (\ref{G1eqt}):
\begin{eqnarray}
\label{tmunuJordanmain}
&& t_\mu {}^\nu = \Omega^{-2} ({T^m}_\mu {}^\nu +
{{T^\varphi}_\mu}^\nu) - {1\over 8\pi G} {G^{(2)}}_\mu {}^\nu 
\nonumber \\
&& \quad \quad + {\Omega^{-2} \over 8\pi G}
\left[\nabla_\mu \nabla^\nu \Omega^2 
- {\delta_\mu}^\nu \Box \Omega^2 \right] \; .
\end{eqnarray}
Comparing this with Eq. (\ref{ttmunu}), we can see the calculation
should be similar to the Einstein frame case, except that
we have an overall conformal factor $\Omega^{-2}$ multiplying 
the matter and scalar field energy-momentum, and that there are
some uniquely Jordan frame contributions to $t_\mu {}^\nu$
(second line). These differences originate from the differences 
in the metric equation (\ref{metriceqtJordan}) from the Einstein equation:
a non-constant effective `$G$', and extra sources for $G_\mu {}^\nu$ beyond
the matter and scalar energy-momentum.

Phrased in this way, it is not obvious that the integral of momentum flux
should imply geodesic motion in Jordan frame. In fact, it does not in general.
Relegating the details to Appendix \ref{app:jordan},
it can be shown that the center of mass of an object moves according to:
\begin{eqnarray}
\label{Jordanmain2main}
M \ddot X^i = - M \, \partial_i \Phi_0 + (1 - \epsilon)
\alpha  M \, \partial_i \varphi_0
\end{eqnarray}
where $\Phi_0$ and $\varphi_0$ are respectively the background/environmental (time-time) metric
perturbation and scalar field, $M$ is the inertial mass and $X^i$ is the center of mass
coordinate.
This is manifestly consistent with the Einstein frame equation of motion
Eq. (\ref{Einsteinmain}) once we recognize that
$\tilde \Phi_0 = \Phi_0 - \alpha \varphi_0$ according to Eq. (\ref{smallconformal}).

The parameter $\epsilon$ is controlled by the relative
size of the object's gravitational potential and the environmental scalar field; see
Eqs. (\ref{thinshell}) \& (\ref{nothinshell}).
An unscreened object, which has $\epsilon = 1$, would
move on a geodesic in Jordan frame, just as expected
for an infinitesimal test particle.
A screened object, where the chameleon mechanism operates
to make $\epsilon \sim 0$, would {\it not} move on a geodesic
even in Jordan frame. 
In other words, it appears as if for screened objects, the
gravitational mass and the inertial mass are unequal.
This can be made more explicit if we make a simplifying assumption,
that both $\Phi_0 - \alpha\varphi_0$ and $\varphi_0$ are sourced mainly
by density (Eq. (\ref{sourceenvironment})), in which case
$\varphi_0 = 2\alpha \Phi_0/(1 + 2\alpha^2)$, and the equation of
motion simplifies to
\begin{eqnarray}
\label{Jordanmain3main}
M \ddot X^i  = M \left[ {{1 + 2\epsilon \alpha^2} \over {1 + 2\alpha^2}} \right]
\partial_i \Phi_0
\end{eqnarray}
which is consistent with its Einstein frame counterpart
Eq. (\ref{Einsteinmain2}). Thus, in Jordan frame,
\begin{eqnarray}
\label{twomasses2}
&& {\,\rm inertial \, \, mass \,} = M \nonumber \\
&& {\,\rm grav. \, \, mass\,} = M  \frac{1 + 2 \epsilon
\alpha^2} { 1 + 2 \alpha^2 }  \; .
\end{eqnarray}
An unscreened object is subject to a gravitational force that is $1 + 2\alpha^2$ larger
than a screened object. For theories such as $f(R)$, this is not a small effect: 
$\alpha = 1/\sqrt{6}$ implies $1 + 2 \alpha^2 = 4/3$ (Appendix \ref{app:fR}).
This might come as a surprise in the Jordan frame, where the metric is minimally coupled
to matter (Eq. (\ref{actionJordan})). Despite conventional wisdom,
specifying the Jordan frame metric does not completely specify the motion of an extended 
object. The motion of an object should be ultimately determined by energy-momentum
conservation. Since the scalar field contributes energy-momentum, it retains a {\it direct} influence on the
motion of the object. This direct influence magically 
cancels only for unscreened objects, which
thus move on geodesics.\footnote{The cancellation occurs between the effective
Newton constant $G\Omega^{-2}$ and the 
uniquely Jordan frame terms $\Omega^{-2} \partial \partial \Omega^2$
in the pseudo-energy-momentum (Eq. (\ref{tmunuJordanmain})).}
The key physics is the (non-relativistic
and classical) renormalization of scalar charge
by the nonlinear chameleon mechanism, as discussed in \S \ref{einstein}. 
An object's coupling to the background scalar field is suppressed if it is screened,
and unsuppressed if it is not. 
\footnote{For a screened object (let's say a galaxy halo) in an unscreened environment,
it is worth noting that the scalar field $\varphi$ experiences
a jump at the thin shell of the object (from a large expectation value outside to a small one inside),
which means that while the Einstein frame metric perturbations are smooth across the shell,
the Jordan frame metric perturbations jump (Eq. (\ref{smallconformal})).
This resolves what might perhaps be puzzling to some: that the screened halo as a whole
does not move on the Jordan frame geodesic, while infinitesimal particles inside it do.
The key observation is that they see rather different overall Jordan frame metrics (even after
accounting for the fact that a particle inside the object of course sees additional
metric perturbations arising from other particles of the object).
In Einstein frame, there is no mystery: neither the screened halo nor its constituent particles
experience any scalar force. We thank Wayne Hu for discussions on this issue.
}


\begin{figure*}[htb!]
\begin{center}
\raisebox{.7cm}{\includegraphics[width=5.7cm]{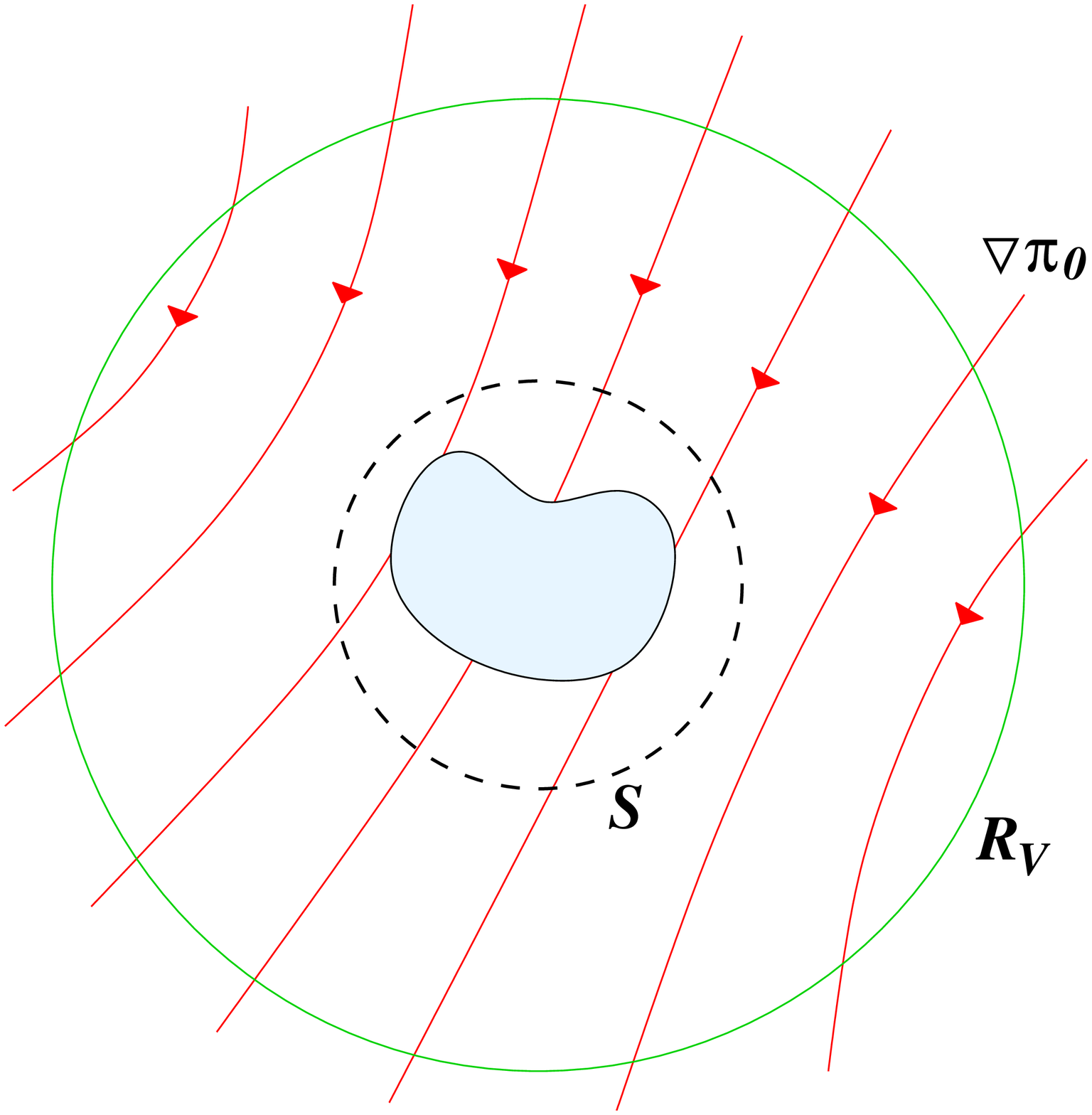}}
\hspace{.7cm}
\raisebox{3.3cm}{\Large $=$}
\hspace{.7cm}
\includegraphics[width=8cm]{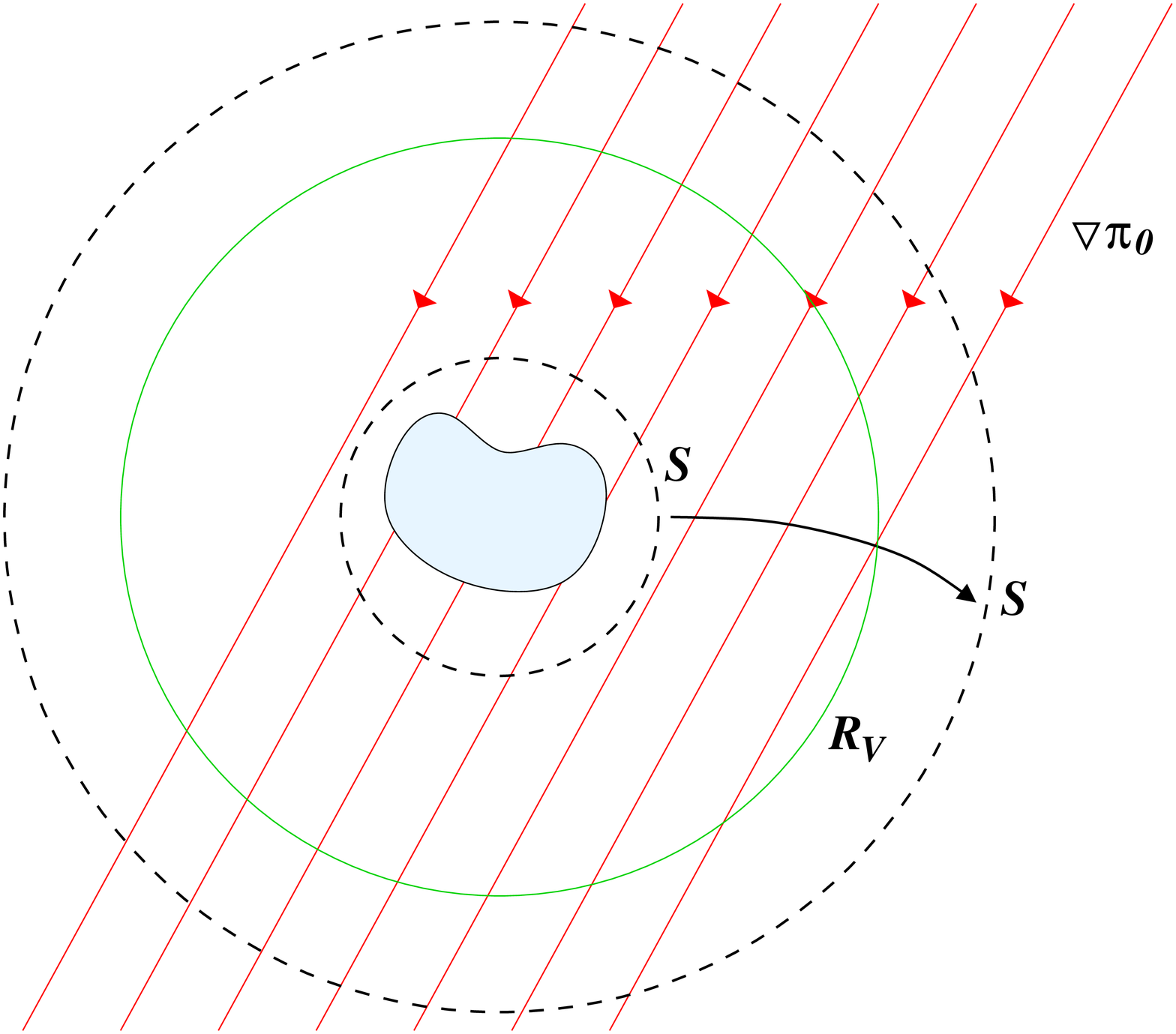}
\end{center}
\caption{\label{trick} Our mathematical trick.}
\end{figure*}

\section{The Problem of Motion in Scalar-Tensor Theories --- Screening by Strong Coupling 
{\it \`a la} Vainshtein}
\label{strongcoupling}

Another class of scalar-tensor theories where the scalar is efficiently screened at short distances is massive gravity theories. These are not scalar-tensor theories to the letter, for at scales of order of the graviton's Compton wavelength, and larger, there is no propagating scalar degree of freedom. However, at much shorter scales the (helicity-zero) longitudinal polarization of the graviton behaves like a scalar universally coupled to matter with gravitational strength, and the dynamics is accurately described by a scalar-tensor theory \cite{AGS}.
What makes these theories potentially interesting is that this effective scalar comes equipped with important derivative self-interactions, which can make it decouple at short distances from compact objects, thus recovering agreement with solar system tests. This is known as the Vainshtein effect, and it shows up explicitly in the DGP model \cite{DGP}---which is a much healthier alternative to the minimal Fierz-Pauli theory of massive gravitons.
However as we will see shortly, unlike the chameleon mechanism the Vainshtein effect does not lead to $O(1)$ violations of the equivalence principle---in fact it leads to the same violations one would have for a purely {\em linear} Brans-Dicke theory, which are
$O(1/c^2)$.

For simplicity and definiteness, we will analyze the example of DGP. Everything we say can be straightforwardly extended to the `galileon', which was argued in ref.~\cite{NRT} to be the broadest generalization of (the scalar-tensor limit of) DGP that robustly implements the Vainshtein effect without introducing ghost degrees of freedom. 
The scalar sector of DGP (in Einstein frame) has the Lagrangian \cite{LPR}
\be \label{DGP}
{\cal L}_\pi = -3 \mpl^2 (\di \pi)^2 - 2 \frac{\mpl^2}{m^2} (\di \pi)^2 \Box \pi +  \pi \, {T^m} {}_\mu {}^{\mu} 
\; ,
\ee
where $m$ is the DGP critical mass scale---the analogue of the graviton's mass---and is generally
set to the current inverse Hubble scale.
We are using a dimensionless normalization for the scalar field $\pi$, so that the 
Jordan-frame metric to which matter fields couple minimally is
\be
h_{\mu\nu} = \tilde h_{\mu\nu} + 2\pi \, \eta_{\mu\nu} \; .
\ee
That is, in the non-relativistic limit $\pi$ is a shift in the Newtonian potential:
\be
\Phi = \tilde \Phi + \pi \; .
\ee

Comparing Eq.~(\ref{DGP}) with Eq.~(\ref{EF}), we can see that
$\pi$ here is equivalent to $\varphi/\sqrt{6}$ there.
We adopt the non-canonically normalized $\pi$ in this section (not to be confused with
the number $3.141...$) to ease comparison with the DGP literature.
In the language of the previous section,
DGP (and its galileon generalization) has $\alpha = 1/\sqrt{6}$, meaning
its Jordan frame action has no quadratic scalar kinetic term. In this respect,
DGP/generalizations resemble $f(R)$
or Brans-Dicke theory with $\omega_{\rm BD} = 0$
(Appendix \ref{app:fR}), but the resemblance is superficial because
DGP/generalizations have higher derivative interactions which allow
a completely different screening mechanism.

A compact source creates a $\pi$ profile that scales like $1/r$ at large distances (linear regime).
However in approaching the source the non-linear term\footnote{Non-linear in the $\pi$ equation of motion, {\it cubic} in the action.} 
of Eq.~(\ref{DGP}) starts becoming important, thus changing the behavior of $\pi$ to $\sqrt{r}$. This suppresses the force mediated by $\pi$ for all objects that are inside this non-linear `halo' surrounding the source. The size of this halo is set by the so-called Vainshtein radius,
\be
R_V \equiv \bigg( \frac M {\mpl^2 \,  m^2}\bigg) ^{1/3} \; ,
\ee
where $M$ is the mass of the source, and $m$ is the graviton mass scale
(about Hubble scale today).

Despite the qualitative similarities, this kind of non-linearity differs significantly  from that in the chameleon case. The point is that the strong non-linearities $\pi$ experiences close to the source do not renormalize the source's total scalar charge.
To see this, notice that the source-free part of the Lagrangian (\ref{DGP}) has a shift-symmetry,
\be
\pi \to \pi + a  \; .
\ee
As a consequence, $\pi$'s equation of motion is the divergence of the associated N\"other's current
\be
\di_\mu J^\mu = -{T^m} {}_\mu {}^{\mu} \; ,
\ee
where $J^\mu$ is a non-linear function of first and second derivatives of $\pi$.
We therefore have a non-linear Gauss-like law for $\pi$, whose source is the trace of the matter
energy-momentum tensor.
For a time-independent solution, this equation is readily integrated via Gauss's theorem to yield
\be
\oint_S \vec J \cdot d\vec a = Q_{\rm tot}
\ee
where $S$ is any surface, and $Q_{\rm tot}$ is the total scalar charge
\footnote{$Q_{\rm tot}$ here is proportional to $Q$ in Eq. (\ref{Qdef}). The difference in normalization is related to the different normalizations for $\varphi$ and $\pi$.}
enclosed by $S$, defined as the volume integral of (minus) the matter $T^m {}_\mu {}^{\mu}$---the total mass for non-relativistic sources. So, for instance, if we consider a single compact spherical source at the origin and we take $S$ to be a sphere centered at the origin, with radius much larger than $R_V$, then on $S$ $\pi$ is in the linear regime, and we have
\be \label{gauss}
6 \mpl^2 \oint_S \vec \nabla \pi  \cdot d\vec a \simeq M \quad \Rightarrow \quad \pi \simeq - \frac{1}{3}\frac{ G M}{r}  \qquad r \gg R_V \; ,
\ee
regardless of how non-linear $\pi$'s dynamics may be inside the Vainshtein region.
The mass $M$ is thus also the scalar charge, and 
according to the analysis of \S \ref{QFT}, $M$ then is the coupling between the object and 
very long-wavelength external/background $\pi$'s. 
This implies there is no chameleon-like non-relativistic renormalization of the scalar charge, and
no $O(1)$ violation of the equivalence principle. 
The only violations of the equivalence principle we may have are associated with how 
relativistic the source's internal structure is, i.e.~on how the volume integral of 
$T^m {}_\mu {}^\mu$ differs from the total ADM mass, which is a post-Newtonian $O(1/c^2)$ effect
\footnote{Notice that since in Eq.~(\ref{gauss}) we are considering a huge volume with sizable non-linearities for $\pi$, we have to make sure that $\pi$'s contribution to the total ADM mass be negligible. It is straightforward to check that this is indeed the case.}.
For the benefit of the skeptical reader, in Appendix \ref{forceVainshtein} 
we explicitly check this statement along the lines of the previous sections.

This would be the end of the story, if our universe were made up of very isolated compact sources, with typical distances between one another much bigger that their typical Vainshtein radius. Unfortunately we live precisely in the opposite limit. For macroscopic sources the Vainshtein radius is so huge that treating those sources in isolation makes no sense. For instance, for the sun we have
\be
R_V^\odot \sim 10^{21} \; {\rm cm} \; ,
\ee
much bigger than the distance to the closest stars. We can repeat the same estimate at all scales, and check that nowhere in the universe $\pi$ is in the linear regime.
This makes a comprehensive study of the effects of $\pi$ in our universe a formidable task, because by definition of a non-linear system, we cannot simply superimpose the $\pi$ fields due to different sources. However here we want to focus on possible violations of the equivalence principle, and for these we will be able to draw surprisingly general conclusions.

Consider then the more realistic situation of an object in the presence of other sources. To be as general as possible, let's imagine that the object in question has sizable multiple moments as well. The situation is schematically depicted in Fig.~\ref{trick}(left). We would like to compute the force acting on the object following the approach of the previous sections. In order to approximate the $\pi$ field created by the other sources ($\pi_0$, in the figure) as a linear pure-gradient field, we have to draw our surface $S$ very close to the object. In fact since $\pi$ is deep
in the non-linear regime, we have two problems
\begin{enumerate}
\item How do we separate  $\pi_0$ from $\pi_1$---the $\pi$ generated by the object? The total $\pi$ is not simply the sum of the two contributions.
\item What is the effect of the object's multipole moments? In a non-linear regime, the multipoles of $\pi$ are not directly related to the object's multipoles. Also, their behavior with $r$ w.r.t.~to the monopole's is different than in the linear case.
\end{enumerate}
As to point 1, we are rescued once again by the symmetries of the $\pi $ dynamics, which will also make point 2 uninteresting for our purpose. 

Besides the aforementioned shift-symmetry on $\pi$, the source-free part of the Lagrangian (\ref{DGP}) is invariant under a constant shift in the {\em derivative} of $\pi$,
\be
\di_\mu \pi \to \di_\mu \pi + c_ \mu \; ,
\ee
dubbed `galilean invariance' in ref.~\cite{NRT, AADNR}. In fact this is what makes DGP---and suitable generalizations thereof---capable of the Vainshtein effect in the presence of generic localized sources, and of sustaining interesting non-linear cosmological solutions \cite{NRT}. Galilean invariance allows us to add a pure-gradient field to {\em any} non-linear solution, to get a new non-linear solution. 
Consider therefore the $\pi$ field generated by all other sources {\em in the absence of our object}, and denote this by $\pi_0$.
Now, if our object is much smaller than the typical variation scale of $\pi_0$, which is set by the distance to the other sources, then we can  approximate $\pi_0$ as a constant-gradient field in a neighborhood of the object.
This means that in such a neighborhood, we can freely add $\pi_0$ to the non-linear solution $\pi_1$ we would have for the object in isolation, i.e.~{\em in the absence of the other sources}. The full $\pi$ field
\be \label{split}
\pi = \pi_0 + \pi_1
\ee
is a consistent solution to the non-linear problem in the neighborhood in question.
It is in fact {\em the} solution, because it has {\em (a)} the right source---the object, and {\em (b)} the right boundary conditions: in moving away from the object, the perturbation in $\di_\mu \pi$ induced by the object decays away as $1/\sqrt{r}$ and we are left with $\di_\mu \pi_0$. 

In summary, thanks to Galilean invariance the splitting of Eq.~(\ref{split}) is well defined whenever the object is much smaller than its typical distance to nearby sources, which is a much milder requirement than linearity. We can then draw our sphere $S$ very close to the object, as in Fig.~\ref{trick}(left), and compute the force acting on it along the lines of the previous sections. We do this explicitly in Appendix \ref{forceVainshtein} for a spherical object. However for an irregular object with sizable multipole moments, we do not even know the non-linear solution $\pi_1$ for the object in isolation. Nevertheless, here we can use a mathematical trick. We have argued that the full solution on $S$ is the field the object would have in isolation, $\pi_1$, plus a constant-gradient field, $\pi_0$. Nothing prevents us from considering a physically different situation, but with the same total field on $S$. This would of course yield the same result for the total force acting on the object. A very convenient situation to consider is, the object in isolation with the same background constant-gradient $\pi_0$ as in our case, but extended {\em linearly} throughout space, as depicted in Fig.~\ref{trick}(right). That is, we can forget about the external sources, and declare that in fact $\pi_0$ has a constant gradient throughout space. In doing so, we are not changing the total $\pi$ and its gradient on $S$---thus we are not changing the total force we would compute by integrating over $S$. We can now change the radius of $S$ and bring it outside the Vainshtein region, where $\pi$'s dynamics is linear, its multipoles decay faster than the monopole, and the computation is precisely as for a linear scalar field. By changing the radius of $S$, we are modifying the total mass enclosed by $S$ by $\pi$'s contribution to it. As long as this is negligible, as can be explicitly checked, the total force computed this way does not depend on the radius of $S$.


We therefore conclude that, as far as the coupling of an extended object to a background $\pi$ field is concerned, theories with the Vainshtein effect behave {\em exactly} as the linear Brans-Dicke theory.
The Vainshtein mechanism entails equivalence principle violations only of the post-Newtonian $O(1/c^2)$
type.


\section{Discussion --- Observational Tests}
\label{tests}

Let us briefly summarize the main results so far, especially those relevant for
a discussion of observational tests. 

{\bf 1.} In theories where screening of the scalar field occurs via
strong coupling effects (Vainshtein), such as DGP, the coupling of objects to the external scalar
fields is unsuppressed to good approximation for all objects
whose mass is not dominated by the gravitational binding energy.
They therefore fall at the same rate under gravity. There remains, however, equivalence principle
violations at the level of $\Phi$, or $O(1/c^2)$ in the language of post-Newtonian expansion, 
just as in Brans-Dicke theory, where
$\Phi$ is the gravitational potential depth of the object in question.
Order one violation of the equivalence principle is visible only when comparing
the motion of relativistic objects such as black holes ($|\Phi| \sim 1$) with the motion of 
less compact objects such as stars ($|\Phi| \ll 1$).
We will focus on observational tests of the chameleon mechanism instead, where
the equivalence principle violations are more dramatic and readily detectable
even in non-relativistic/non-compact objects.

{\bf 2.} The chameleon mechanism implies that the coupling of screened
objects to the external/background scalar field is highly suppressed relative to unscreened objects. They therefore fall at
significantly different rates. More precisely, the equation of motion for the center of
mass of an object is, in Jordan frame (Eq. (\ref{Jordanmain2main})):
\begin{eqnarray}
\label{testsEOM}
M \ddot X^i = - M \partial_i \Phi_0 + (1 - \epsilon) \alpha M \partial_i \varphi_0\end{eqnarray}
where $X^i$ is the center of mass coordinate, $M$ is the inertial mass,
$\Phi_0$ is the time-time part of the external/background metric and
$\varphi_0$ is the external/background scalar field (through which the object moves).
Here we have adopted a convenient parameterization of the relevant conformal factor $\Omega^2 (\varphi)$
in the scalar-tensor action (Eqs. (\ref{actionJordan}) \& 
(\ref{alphaphi})):
\begin{eqnarray}
\label{testsOmega}
\Omega^2 (\varphi) = 1 - 2\alpha \varphi \quad , \quad |\alpha \varphi| \ll 1
\end{eqnarray}
where $\alpha$ is a constant. Eq. (\ref{testsOmega}) can be thought of as a
small field expansion  of e.g.~$\exp [-2\alpha\varphi]$.
The quantity $\epsilon$ represents the degree of screening.
Unscreened objects have $\epsilon = 1$, while screened objects have $\epsilon < 1$.
The screening parameter $\epsilon$ is controlled by 
(Eqs. (\ref{thinshell}) \& (\ref{nothinshell})):
\begin{eqnarray}
\label{testsEpsilon}
\epsilon \simeq {\varphi_*\over 2 \alpha (GM/r_c)} \quad \mbox{if} \quad
{\varphi_* \over 2 \alpha (GM/r_c)} < 1  \quad \mbox{(screened)} &&\nonumber \\
\epsilon \simeq 1 \quad \mbox{if} \quad
{\varphi_* \over 2 \alpha (GM/r_c)} \gsim 1 \quad \mbox{(unscreened)}&&
\end{eqnarray}
where $r_c$ is the size of the object in question, and
$\varphi_*$ is the value of the external scalar field.
Screening is therefore controlled by the ratio of
the exterior scalar field and the gravitational potential of the object.
Unless the values of the two are finely tuned, one generally
expects $\epsilon \ll 1$ if screening occurs at all.
Eq. (\ref{testsEOM}) tells us only unscreened objects, those with
a sufficiently shallow gravitational potential,
have equal inertial mass and gravitational mass in Jordan frame.

{\bf 3.} The precise ratio of inertial mass to gravitational mass
predicted by Eq. (\ref{testsEOM}) depends on how the external scalar field $\varphi_0$ 
is sourced. There are two useful limits to consider. One is when the
external scalar field is sourced primarily by density (Eq. (\ref{sourceenvironmentJordan}))
in which case the equation of motion is well-approximated by
\begin{eqnarray}
\label{testsEOM2}
M \ddot X^i = - M \left[ {1 + 2 \epsilon \alpha^2 \over 1+ 2\alpha^2 } \right]
\partial_i \Phi_0
\end{eqnarray}
which implies the unscreened and screened gravitational mass are related by
\begin{eqnarray}
\label{testsMass}
\mbox{unscreened g.~mass} = (1 + 2\alpha^2) \times
\mbox{screened g.~mass}
\end{eqnarray}
The other limiting case is when the external scalar field profile is
determined primarily by the potential, in which case $\varphi_0$
is Yukawa suppressed on sufficiently large scales (scales larger than
the Compton wavelength of the scalar field $\sim 1/m_\varphi$).
If $\varphi_0$ is Yukawa suppressed, the second term on the right
of Eq. (\ref{testsEOM}) is small regardless of whether the object
is screened or not. In this regime, one cannot observe equivalence
principle violations by comparing motions of different objects.
As we will discuss below, it is therefore important to avoid Yukawa suppression
when performing observational tests.

{\bf 4.} Observationally, the single most important parameter besides
$\alpha^2$ in
a chameleon theory is the quantity $\varphi_*/\alpha$
(see Eq. (\ref{testsOmega})). 
For a given object, the relevant $\varphi_*$ is the scalar field
value in its immediate surrounding. If this is sufficiently small,
the object will be screened according to Eq. (\ref{testsEpsilon}).
In other words, if the object is sitting in some high density environment
which is itself screened (i.e. small scalar field value), it is likely that
the object is also screened. We will refer to this as blanket screening.
In an environment where blanket screening of all objects applies,
violations of equivalence principle will be hard to detect.

\begin{figure}[htb!]
\begin{center}
\includegraphics[width=7.5cm]{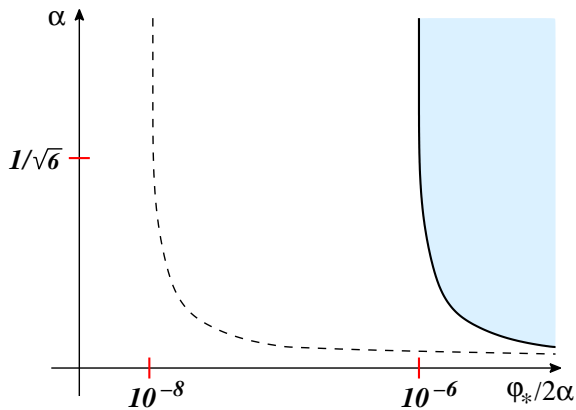}
\end{center}
\caption{\label{bounds} Schematic experimental bounds in the chameleon parameter space. The shaded region is already excluded, by demanding that the Milky way be screened. The dashed line is the improvement we propose. Further improvement is possible using Lyman-alpha clouds. For $\alpha \ll 1$ the chameleon is very weakly coupled to matter to begin with, and clearly this relaxes the bound on the other parameter. Curvature invariant theories such as
$f(R)$ have $\alpha = 1/\sqrt{6}$, though this value is not protected by 
symmetry. The generic expectation is $\alpha \sim O(1)$.}
\end{figure}

Conversely, if the object is sitting in an environment where the density
is at the cosmic mean or even lower (voids), the chances are better that
the object is unscreened because the relevant $\varphi_*$ is larger.
In other words, to find unscreened objects, it is useful to 
avoid overdense regions and to focus on objects with a shallow
gravitational potential such that (Eq. (\ref{testsEpsilon})):
\begin{eqnarray}
\label{testsUnscreen}
{GM \over r_c} \lsim {\varphi_* \over 2 \alpha}
\end{eqnarray}
where $\varphi_*$ is taken to be the cosmic mean value.
We already know ${\varphi_*/(2\alpha)} \lsim 10^{-6}$ from requiring that the Milky Way
be screened to avoid gross deviations from GR in solar system
experiments \cite{Hu:2007nk} (see also \cite{Reid} for a recent revision of
the Milky Way's mass).
This leaves us an interesting window for testing
theories like $f(R)$. The smallest galaxies of interest have
$GM/r_c \sim 10^{-8} (v/30 {\,\rm km/s})^2$. Therefore, if
$\varphi_* /\alpha$ for our modified gravity theory satisfies
\begin{eqnarray}
\label{testsRange}
10^{-8} \lsim {\varphi_* \over 2\alpha} \lsim 10^{-6} \;\;
\mbox{(observationally interesting)} 
\end{eqnarray}
massive galaxies at the high end would
be screened while less massive galaxies at the low end would not,
offering many opportunities for observing equivalence principle violations,
as we will see. The parameter space for chameleon mechanism is illustrated in Fig. \ref{bounds}.

{\bf 5.} To make 
our discussion more concrete, it is useful
to keep in mind a particular class of examples, the so called $f(R)$ theories.
As shown in Appendix \ref{app:fR}, $f(R)$ corresponds to a specific
value for $\alpha$:
\begin{eqnarray}
\label{testsAlphafR}
\alpha = 1/\sqrt{6} \quad {\rm for} \,\, f(R)
\end{eqnarray}
and so the unscreened gravitational force is $4/3$ higher than
the screened one (Eq. (\ref{testsEOM2})). This value for $\alpha$
arises simply from demanding that $\varphi$ be an auxiliary field in Jordan frame
(i.e.~$h(\varphi) = 0$ in Eq. (\ref{actionJordan})).
It is worth stressing however that from a theoretical viewpoint $f(R)$ theories are not favored over more generic scalar-theories. For instance the property that $\varphi$ has no kinetic energy in Jordan frame is not protected by any symmetry, thus one generically expects that upon quantum corrections an $f(R)$ theory will become a scalar-tensor theory not belonging to the $f(R)$ class.
In fact, the resulting $\alpha$ could be even {\it larger} than the
$1/\sqrt{6}$.
On the other hand, the structure of the $\pi$ Lagrangian in DGP and in galileon theories {\em is} protected by a symmetry---Galilean symmetry---which forbids the quantum generation of Lagrangian terms with fewer than two derivatives per field \cite{LPR}.

The most important parameter in an $f(R)$ theory is 
$f_R \equiv df/dR$ evaluated at the cosmic mean, and it can be shown
\begin{eqnarray}
f_R = - 2 \alpha \varphi_* \; .
\end{eqnarray}
Appendix \ref{app:fR} also shows that common parameterizations of $f(R)$ theories
connect $\varphi_*$ and the Compton wavelength $1/m_\varphi$ of the scalar field:
\begin{eqnarray}
\label{testsComptonfR}
m_\varphi^{-1} \sim \sqrt{\varphi_* \over \alpha} H^{-1} 
\end{eqnarray}
where $H^{-1}$ is the Hubble scale. The fifth force is Yukawa
suppressed between a pair of objects separated by scales larger
than $1/m_\varphi$. This places an important restriction on
the range of observational tests available for these theories.
One should keep in mind though that Eq. (\ref{testsComptonfR})
can be evaded by considering more general potentials than is commonly
discussed (see Appendix \ref{app:fR}).

\begin{figure}[htb!]
\begin{center}
\includegraphics[width=8.5cm]{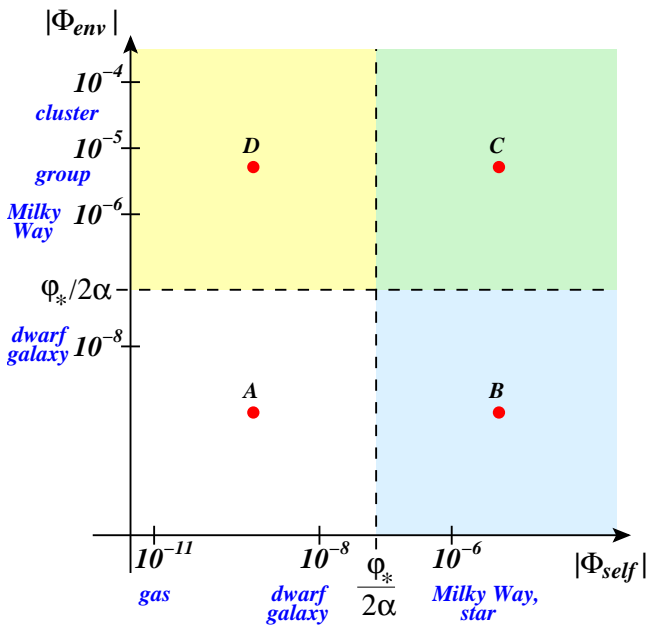}
\end{center}
\caption{\label{comparison} A schematic illustration of observational tests.
$\Phi_{\rm self}$ and $\Phi_{\rm env}$ represent the gravitational potentials
of an object and its environment. They can be thought of as $\sim (v/c)^2$, where
$v$ is the internal velocity. The value $\varphi_*/(2\alpha)$ delineates screening
or lack thereof---objects/environments with a potential deeper than $\varphi_*/(2\alpha)$ are
screened (shaded), and those with a shallower potential are unscreened (unshaded).
Current constraints tell us $\varphi_*/(2\alpha)$ has to be less than $\sim 10^{-6}$.
There are many comparison tests one can make. 
For instance, an unscreened diffuse gas cloud residing
in a dwarf galaxy (A), versus a screened star residing in the same
galaxy (B)---A falls faster than B. This situation can be replicated on a larger scale
e.g. a dwarf galaxy in the fields/voids (A) versus a massive galaxy in the same
fields/voids (B). Another example: a dwarf galaxy
out in the fields/voids (A), versus a dwarf galaxy residing in a 
group or cluster (D)---D is blanket screened by its environment and would
exhibit no equivalence principle violations in its internal motions of
gas clouds and stars, while A would have observable violations.
On the other hand, a massive galaxy would have no such (internal) equivalence principle
violations whether it be in the fields/voids (B) or in a group/cluster (C).
}
\end{figure}

{\bf Observational tests.} Below we list a series of possible observational tests \cite{othertests}.
They can be viewed in one of two ways: if they turn up positive, we will have a measurement
of $\varphi_*/\alpha$ and $\alpha$ 
within the chameleon framework and a refutation of screening by strong coupling; 
if they turn up negative, we can interpret the results either as an upper limit on $\varphi_*/\alpha$
and $\alpha$ 
for the chameleon mechanism or as indirect support for the strong coupling picture (or as yet another confirmation of GR of course).
Tests 1 and 2 focus on the bulk motion of individual galaxies while
tests 3, 4 and 5 focus on the internal motion of their constituents.
Fig. \ref{comparison} illustrates these
various  possibilities. Further tests are discussed at the end.

{\bf Test 1.} Eq. (\ref{testsEOM2}) tells us that
unscreened galaxies (whose gravitational potential is sufficiently
shallow that Eq. (\ref{testsUnscreen}) is satisfied) accelerate faster
compared to screened galaxies by a factor of $1 + 2\alpha^2$, which is
$4/3$ for $f(R)$ theories. We therefore expect small galaxies to move faster
than large ones in general. One must be careful not to confuse this
with dynamical friction which could cause a similar effect 
(sometimes referred to as velocity bias; see \cite{carlberg,klypinv}).

A common way to probe the peculiar motion of galaxies is via
redshift space distortions. On large (linear) scales, where dynamical friction
is unimportant, one could use the anisotropy of the redshift space correlation
function to measure $\beta = b^{-1} d{\,\rm ln\,} D/ d{\,\rm ln\,}a$ where
$b$ is the linear galaxy bias, $D$ is the linear growth factor and 
$a$ is the scale factor \cite{kaiser,hamilton}.
The new twist introduced by the chameleon mechanism is that
effectively $d{\,\rm ln\,} D/d{\,\rm ln\,} a$, which arises from
the ratio between peculiar motion and mass overdensity, could be different for
different size galaxies. This means if one takes the ratio $\beta_1/\beta_2$ between
small and large galaxies, and multiplies it by the appropriate ratio
of the real space power spectrum $P$
\footnote{This can be obtained from looking at Fourier modes that point
transverse to the line of sight.}
to remove the dependence on galaxy bias, GR would return unity while
the chameleon mechanism would not i.e.
\begin{eqnarray}
\label{testsBeta}
{\beta_1 \over \beta_2} \sqrt{P_1 \over P_2} &=& 1 \quad {\,\rm for \,\, GR} \nonumber \\
&\ne& 1 \quad {\, \rm for \,\, chameleon}
\end{eqnarray}
where $1$ denotes the unscreened (small) galaxies and $2$ denotes the 
screened (large) galaxies. The precise prediction for chameleon theory 
is a bit complex and will be presented elsewhere. 
This is a worthwhile test to carry out---if anything, it serves as
a consistency test for GR---but as a test of $f(R)$ theories, one must
take care to avoid Yukawa suppression.
For those classes of $f(R)$ theories that obey Eq. (\ref{testsComptonfR}),
the Compton wavelength is fairly short given current upper limits on $\varphi_*/\alpha$
(Eq. (\ref{testsRange})). In that case, Yukawa suppression implies the relative motion between
pairs of galaxies separated on linear scales is likely insensitive
to the scalar force.
We emphasize, however, that the relation between Compton wavelength and $\varphi_*/\alpha$ 
in Eq. (\ref{testsComptonfR}) is not general and can be relaxed.

Alternatively, one could focus on the relative peculiar motion between galaxies
on small scales, but then one needs to contend with the possibility
of both blanket screening and dynamical friction.
One way around these problems is to adopt the method of \cite{strauss} which allows one to
study the pairwise velocity dispersion as a function of environment.
Voids would be the ideal environment which bypasses both problems, and allows
one in principle to observe the systematically different peculiar velocities
between screened and unscreened galaxies.
However, because the observed peculiar motion is always galaxy weighted,
one must devise a way to disentangle apparent velocity difference due to different
(nonlinear) galaxy biases from genuine velocity difference due to
the chameleon mechanism.

In summary, the cleanest galaxy bulk motion test is probably to examine
redshift distortions on large (linear) scales where there
are no complications from dynamical friction or nonlinear galaxy bias.
In this case,
a necessary requirement for order unity observable effects is that
the scalar Compton wavelength (at cosmic mean density) cannot be too small,
i.e. it must extend to linear scales.

{\bf Test 2.} In environments such as voids, where blanket screening of all galaxies does
not occur, the systematic difference in the motion of small versus large galaxies
implies their density distributions would differ from standard expectations.
For instance, galaxies are expected to stream away from voids. The fact that small galaxies
stream at a faster velocity means voids defined by them would appear larger than
expectations based on the large galaxies.
This prediction of the chameleon mechanism is intriguing because 
as emphasized by Peebles \cite{peeblesvoid} the observed voids defined by the small galaxies appear
too large compared to predictions from $\Lambda$CDM 
model, whereas the observed void sizes defined by larger galaxies appear consistent
with the standard model \cite{tinker,tikhonov}.
For the chameleon mechanism to provide a viable solution to the void problem,
one must consider models that evade the Compton wavelength condition
of Eq. (\ref{testsComptonfR}).

More generally, because screened and unscreened galaxies move differently, their
clustering bias will evolve differently, giving potentially observable effects
which will be explored elsewhere (see e.g.\cite{HP}).

{\bf Test 3.} Perhaps the cleanest way to test the chameleon mechanism is to
examine the internal motions of objects inside an unscreened galaxy i.e. whose
potential is sufficiently shallow that $GM/r_c \lsim \varphi_*/(2\alpha)$ 
(Eq. (\ref{testsUnscreen})).
The sun has
\begin{eqnarray}
\label{testsStar}
{GM \over r_c} \sim 2 \times 10^{-6} 
\end{eqnarray}
and main sequence stars are expected to have a similar gravitational potential \cite{kippenhahn}.
Current constraints ($\varphi_*/(2\alpha) \lsim 10^{-6}$; Eq. (\ref{testsRange})) 
therefore likely imply the majority of stars 
inside an {\it unscreened} galaxy are themselves {\it screened}, 
because the stars have a fairly deep potential.
On the other hand, diffuse gas inside the same galaxy likely remains {\it unscreened}
because the gravitational potential at a gas cloud should be dominated
by the potential of the galaxy rather than that of the gas cloud itself. For instance
HI gas clouds typically have an atomic number density of $\sim 10 {\,\rm cm^{-3}}$
and a size of $\sim 10 {\,\rm pc}$ \cite{HeilesTroland} giving a tiny self-potential
of $GM/r_c \sim 10^{-11}$. 
\footnote{One might also worry that a galactic disk could have a deeper gravitational
potential than the halo and potentially blanket screen all objects in it.
In practice, the gravitational potential of a disk is comparable to that of the halo
\cite{rhee}.}
Therefore,
at the same radius from galactic center, the HI gas clouds should move with an acceleration 
that is larger than that of the stars by a factor of $1 + 2\alpha^2$, which is $4/3$ for
$f(R)$. Mass estimates constructed from these two tracers will differ
by the same factor
\footnote{When comparing HI rotational velocity with stellar rotational velocity
from a disk galaxy, care should be taken to account for a possible asymmetric drift \cite{BT}.
A proper dynamical modeling should unveil systematic
difference in the mass estimated from HI and that from stars, should the chameleon mechanism
be at work.
}.
By focusing on internal motions, this test avoids issues to do with Yukawa suppression
that plagues the bulk motion tests of some $f(R)$ models i.e. the scalar 
Compton wavelength probably exceeds the size of a galaxy (Eq. (\ref{testsComptonfR})).
To maximize the chance that the galaxy studied is unscreened, one should look for
the smallest galaxies, preferably in voids or at least in the fields
\footnote{Screened galaxies are not interesting for this test because both the stars and
the HI gas would be blanket screened.}.
A small galaxy with internal $v \sim 30 {\,\rm km/s}$ would provide us
either a positive detection of the chameleon mechanism, or an upper limit on $\varphi_*/(2\alpha)$
of about $10^{-8}$, for an $O(1)$ $\alpha$ like in $f(R)$.

{\bf Test 4.} In addition to estimating masses of galaxies from their internal dynamics, 
one could also estimate their masses using gravitational lensing. Photons should behave as
unscreened particles and move on null geodesics in the Jordan frame.
They therefore see the $(\Phi + \Psi)/2$ potential. Unscreened non-relativistic objects
(such as HI gas clouds) on the other hand should move according to the $\Phi$ potential,
while screened non-relativistic objects (such as stars) should move according to
Eq. (\ref{testsEOM}) with $\epsilon \sim 0$. 
Eq. (\ref{metriceqtJordan3}) together with the assumption of negligible Yukawa suppression
then tells us that the mass estimate from photons should equal the mass estimate
from stars, and both should be smaller than the mass estimate from HI by a factor of
$1 + 2\alpha^2$. Carrying out this test might be a bit challenging as one would ideally
like to measure the lensing mass for the smallest galaxies possible to avoid screening
of the whole galaxy. Stacking small galaxies and performing a weak lensing shear measurement
is likely the way to go.

Incidentally, returning briefly to the subject of bulk motion: for the same reasons
articulated above, weak lensing measurements on large scales should yield
consistent results when compared against redshift distortion measurements
of screened (large) galaxies, but inconsistent results when compared against
redshift distortions of unscreened (small) galaxies
\footnote{Modulo Yukawa suppression, as usual.}.
This adds a possible new twist to existing tests of this sort proposed in the literature 
\cite{pengjie}.

{\bf Test 5.} Kesden and Kamionkowski \cite{Kesden} recently proposed an interesting
test of the equivalence principle: if dark matter particles and stars fall at different
rates under gravity, the tidal streams from infalling galactic satellites would be asymmetric.
To this the chameleon mechanism adds a new twist: only unscreened (small) galaxies
will exhibit this asymmetry; screened (large) galaxies will not, because both dark
matter and stars are blanket screened. The current limit 
of $\varphi_*/(2\alpha) \lsim 10^{-6}$ guarantees that the Milky Way does not
exhibit the Kesden-Kamionkowski effect. To see it, one would have to examine smaller
galaxies, preferably out in the fields or voids.
It is also important to emphasize the difference in the origin of equivalence principle
violation in this paper from that in most of the cosmology literature, including \cite{Kesden}.
In most of the literature, equivalence principle violation is there from the beginning, in the sense
that it exists at the level of the microscopic action i.e.~elementary baryons and dark
matter particles are coupled to the scalar differently (see discussion in \S \ref{intro}).
In this paper, there is no explicit equivalence principle violation at the level
of the microscopic action---it gets turned on by the chameleon mechanism and differentiates
not so much between baryons and dark matter, but between objects with a deep gravitational
potential and those with a shallow potential.

The tidal stream test brings to mind an interesting question: if
stars and dark matter inside an unscreened galaxy fall at different rates under
some external gravitational field, shouldn't the galaxy fall apart in the
sense that the stars and dark matter separate (even in the absence
of significant tidal forces)? This should happen at some level, but
should not be a large effect as long as internal gravitational forces dominate
over external ones.

Perhaps the ultimate test objects are low column density 
Lyman-alpha clouds in the intergalactic medium. They have very shallow gravitational potentials
and can therefore be used to probe a much lower $\varphi_*/(2\alpha)$ than are afforded by
dwarf galaxies. Comparing the motions of clouds \cite{Lyalpha1,Lyalpha2}
and galaxies in similar environments could
unveil the chameleon mechanism at work or put severe limits beyond those depicted in Fig. \ref{bounds}.
This is probably a good place to end this discussion. Many of the ideas presented above need to be
worked out in detail, and undoubtedly there are many other possible observational tests.
We hope to pursue these in the future.

\acknowledgments

We thank Kurt Hinterbichler, 
Justin Khoury, Janna Levin and Jacqueline van Gorkum for discussions, 
and Wayne Hu for useful comments.
LH thanks Chris Hirata for raising the issue of the motion of galaxies
in $f(R)$ theories at a CITA workshop in June 2008.
Research for this work is supported by the DOE, grant DE-FG02-92-ER40699,
and the Initiatives in Science and Engineering Program
at Columbia University. 

\appendix


\section{The effect of higher multipoles}
\label{app:multipoles}

We want to show that as long as the external fields $\Phi_0, \Psi_0$ are approximately linear on the spherical surface $S$, higher multipole moments of the object have no effect---as expected, tidal effects are there only if the external field gradients vary appreciably over the object's volume. We will do this explicitly for GR. The generalization to the scalar-tensor case is straightforward in Einstein frame. Our discussion here thus makes clear the surface $S$ can in fact be chosen quite close to the object; it is not necessary to choose a large radius $r$ so that only the monopole survives. As we will see, even the spherical shape of the surface is unnecessary. So, $S$ can just be the object's surface.

Let us not assume that the object's multipole moments are negligible. We have to dot Eq.~(\ref{G2ij}) with $dS_j$, and integrate over a surface $S$ surrounding the object. We need not assume this surface is spherical. The first simplification is that (Eq. (\ref{Einsteinframe1orderB}))
\be
\Phi = \Psi \; ,
\ee
which means that the term proportional to $G^{(1)} {}_i {}^j$ vanishes, and that many terms are equal. Also
\be
\nabla^2 \Phi = 0 \; ,
\ee
as long as the environmental $\rho$ right at $S$ is negligible (which it must be if we are approximating $\Phi_0 = \Psi_0$ as linear). We are left with
\be
G^{(2)} {}_i {}^j = 4 \Phi \, \di_i \di_j \Phi + 2 \di_i\Phi \,  \di_j \Phi - 3 \delta_{ij} (\di_k \Phi)^2 \; .
\ee
The second simplification comes from the fact that we have to integrate this in $dS_j$, over a closed surface. This means that we can add a curl for free. We  use this freedom to eliminate the second-derivative term, thus getting
\be
G^{(2)} {}_i {}^j \to - 2 \di_i\Phi \,  \di_j \Phi + \delta_{ij} (\di_k \Phi)^2 \; .
\ee
The total force thus is
\be \label{total_force}
\dot {\vec  P} = -\frac{1}{8\pi G} \oint _S \big[ 2 \vec E (\vec E \cdot \hat n) - \hat n E^2 \big] dS \; ,
\ee
where we defined the `electric field'
\be
\vec E \equiv \vec \nabla \Phi \; ,
\ee
and $\hat n$ is the outgoing normal. We now decompose $\vec E$:
\be
\vec E = \vec  E_0 + \vec E_1 \; ,
\ee
where the first term is the environmental background, 
\be
\vec E_0 \simeq {\rm const} \; ,
\ee
and the second is due to the object itself.
Expanding Eq.~(\ref{total_force}) we get
\bea \label{decompose_total_force}
-(8\pi G)  \!\! & \dot{\vec  P} & =  \oint _S \big[ 2 \vec E_0 (\vec E_0 \cdot \hat n) - \hat n E_0^2 \big] dS \nonumber \\
	& + & 2 \oint _S \big[  \vec E_0 (\vec E_1 \cdot \hat n) + \vec E_1 (\vec E_0 \cdot \hat n) - \hat n \vec E_0 \cdot \vec E_1 \big] dS \nonumber \\
	& + &\oint _S \big[ 2 \vec E_1 (\vec E_1 \cdot \hat n) - \hat n E_1^2 \big] dS   \; .
\eea
The first line vanishes for constant $\vec E_0$ because
\be
 \oint _S \hat n^j dS = 0 \; ,
\ee
for any closed surface. To show that the last line also vanishes in general, we make use of  a vector form of the Kirchoff integral (see e.g.~ref.~\cite{jackson}): if $\vec E$ is an electrostatic field with no source on $S$, and $G(\vec x - \vec x \, ')$ is a Green's function,
\be
\nabla^2 G(\vec x - \vec x') =  \delta^{3}(\vec x - \vec x \, ') \; ,
\ee
then
\be
\oint _S \Big[ 2 \vec E (\vec \nabla G \cdot \hat n) - \hat n \vec E \cdot \vec \nabla G  
+ \hat n \times \big( \vec \nabla G \times \vec E \big) \Big] dS = 0 \; ,
\ee
whenever the point $\vec x \, '$ is {\em inside} the volume bounded by $S$. To be totally clear, $\vec E$ is a function of $\vec x$, over which we are integrating. $G$ instead is a function both of $\vec x$ and of $\vec x \, '$.
The gradients are taken w.r.t.~$\vec x$. If we replace $\vec E \to \vec E_1$, we multiply by the object's density $\rho_1(\vec x \, ')$, and integrate in $\vec x \, '$ over the object's volume, the l.h.s.~reduces
precisely to the last line of eq.~(\ref{decompose_total_force}). Physically, its vanishing means that the object does not exert any net force on itself---as expected.

We are thus left with the second line of Eq.~(\ref{decompose_total_force})---the mixed terms. Factoring $\vec E_0$ out of the integral, we have to compute the tensor integral
\be
I^{ij} \equiv \oint _S E_1^i \, \hat r^j \, dS = \oint _S \di_i \Phi_1 \,  \hat r^j \, dS  \; ,
\ee
and then take suitable contractions:
\be \label{contractions}
\dot P_k = -\frac{1}{4\pi G}  I^{ij} \big[ \delta^{ij} E_0^k+ \delta^{ik} E_0^j - \delta^{jk} E_0^i \big] \; .
\ee
However, although not manifestly, $I^{ij}$ is a symmetric tensor:
\be
\oint _S \di_i \Phi_1 \,  \hat r^j \, dS  =   \int _V \di_i \di_j \Phi_1 \,  d^3 x  \; .
\ee 
This means that only the first term in brackets in Eq.~(\ref{contractions}) survives, and we have
\be
\dot {\vec  P} = -\frac{1}{4\pi G} \vec E_0 \oint_S \vec \nabla \Phi_1 \cdot d \vec S \; .
\ee
The flux of $\vec \nabla \Phi_1$  through $S$  {\em defines} the total mass enclosed by $S$, and we finally get
\be
\dot {\vec  P} = - M \, \vec \nabla \Phi_0 \; ,
\ee
independently of higher multipole moments of the object's mass distribution.


\section{Mapping to Brans-Dicke and $f(R)$}
\label{app:fR}

It is useful to relate the Jordan frame scalar-tensor action of Eq. (\ref{actionJordan})
to certain classes of scalar-tensor theories in the literature.

Brans-Dicke theory, as it is usually written, corresponds to choosing 
the potential $V = 0$, and defining
\begin{eqnarray}
\phi_{\rm BD} \equiv {1\over G} \Omega^2 (\varphi)
= {1\over G} {\,\rm exp} \left[-{2\alpha\varphi}\right] \; .
\end{eqnarray}
Eq. (\ref{actionJordan}) then becomes
\begin{eqnarray}
S & =&  \int d^4 x \sqrt{-g} {1\over 16\pi} 
\left[ \phi_{\rm BD} R - {\omega_{\rm BD} \over \phi_{\rm BD}} \nabla_\mu \phi_{\rm BD}
\nabla^\mu \phi_{\rm BD} \right] \nonumber \\
&+ &  \int d^4 x \, {\cal L}_m (\psi_m, g_{\mu\nu})
\end{eqnarray}
where
\begin{eqnarray}
\omega_{\rm BD} \equiv {1 - 6\alpha^2 \over 4\alpha^2} \; .
\end{eqnarray}

Phenomenologically allowed Brans-Dicke theory has a large $\omega_{\rm BD}$
which corresponds to a small $\alpha$. With the addition of a suitable potential,
the chameleon mechanism makes even small $\omega_{\rm BD}$ theories
observationally viable. Indeed, $f(R)$ theories correspond to $\omega_{\rm BD} = 0$,
or $\alpha = 1/\sqrt{6}$.
More concretely, consider a scalar-tensor theory with the following
action parameters (Eq. (\ref{actionJordan})):
\begin{eqnarray}
&& \Omega^2 = 1 + {d f \over d \phi_C} \nonumber \\
&& \Omega^4 V = {1\over 2} \left[ \phi_C {d f \over d \phi_C} - f \right]
\end{eqnarray}
where $f$ is some function of a field $\phi_C$, which is related to our familiar $\varphi$ through
\begin{eqnarray}
\label{phiCvarphi}
\Omega^2 = {\,\rm exp\,} \left[- {2\alpha \varphi } \right] = 1 + {d f \over d \phi_C}
\end{eqnarray}
with $\alpha$ set to $1/\sqrt{6}$. Chiba \cite{Chiba1} showed that this
is exactly equivalent to an $f(R)$ theory with
\begin{eqnarray}
&& S = \int d^4 x \sqrt{-g} {1 \over 16 \pi G} [ R + f(R) ] \nonumber \\
&& \quad \quad + \int d^4 x  \, {\cal L}_m (\psi_m, g_{\mu\nu}) \; .
\end{eqnarray}
This can be easily seen by integrating out $\phi_C$, or equivalently $\varphi$, which is a pure auxiliary field.
Eq. (\ref{phiCvarphi}) makes clear why in $f(R)$ theory, $df/dR$, which is simply $df/d\phi_C$ for $\phi_C = R$,
plays the role of a scalar field---for practical purposes, $\Omega^2 \simeq 1 - 2\alpha \varphi$, and
therefore
\begin{eqnarray}
{df \over dR} \simeq - 2 \alpha \,  \varphi
\end{eqnarray}
which we also refer to as $-2 q$ in Appendix \ref{app:jordan} (Eq. (\ref{qdefnew})).

The Compton wavelength $m_\varphi^{-1}$ can be derived from
\begin{eqnarray}
m_\varphi^2 = {\partial^2 V \over \partial \varphi^2} \; .
\end{eqnarray}
Note our somewhat unusual normalizations for $V$ and $\varphi$ in
Eq. (\ref{EF}): the actual potential is $\mpl^2 V$ and the canonically
normalized scalar field is $\mpl \varphi$. These extra factors of $\mpl$
nicely cancel in the definition for the scalar mass $m_\varphi$.
For certain potentials, such as a power law, one can approximate
\begin{eqnarray}
\label{Vassume}
{\partial^2 V \over \partial \varphi^2} \sim {1\over \varphi} 
{\partial V \over \partial \varphi} \; .
\end{eqnarray}
Assuming $\varphi$ is sitting in equilibrium, i.e. the right hand side of
Eq. (\ref{scalarEinstein}) vanishes, we obtain
\begin{eqnarray}
\label{mvarphi}
m_\varphi^2 \sim {\alpha H^2 \over \varphi_*} 
\end{eqnarray}
where we have assumed the local density $\tilde \rho$ is at the cosmic mean,
and $\varphi_*$ is the corresponding scalar field value.
The Compton wavelength in other environments can be computed in an analogous fashion.

We should emphasize that Eqs. (\ref{Vassume}) and thus (\ref{mvarphi}) are by no
means forced upon us. A common parameterization of the $f(R)$ potential does have this form
\cite{Hu:2007nk}, but it is not necessary. For instance, an exponential potential
of the form $V \propto \exp \beta/\varphi$ 
can evade these restrictions, if a suitable choice of the coefficient $\beta$ is made.


\section{Jordan Frame Derivation --- Screening by Chameleon}
\label{app:jordan}

Here, we supply details that are skipped in \S \ref{jordan} for our
Jordan frame derivation of the equation of motion.
It will prove convenient to define a parameter $q$:
\begin{eqnarray}
\label{qdefnew}
\Omega^2 (\varphi) \equiv 1 - 2 q(\varphi) {\quad \rm i.\,e. \quad}
q = \alpha\varphi
\end{eqnarray}
where $\Omega^2(\varphi)$ is the conformal transformation between
Einstein and Jordan frames (see Eqs. (\ref{conformaltransf}) \& (\ref{alphaphi})). 
We assume $|q| \ll 1$.

In addition to the Jordan frame metric equation (\ref{metriceqtJordan}), we
need the scalar field equation:
\be
\label{scalareqtJordan}
h \Box \varphi + \frac12 {\partial h \over \partial \varphi}
\nabla_\alpha \varphi \nabla^\alpha \varphi - {\partial \Omega^4 V \over
\partial \varphi} + \frac12 {\partial \Omega^2 \over \partial \varphi} R  = 0 \; .
\ee

Both the metric equation and the scalar field equation look considerably
more complicated than their Einstein frame counterparts (Eqs. (\ref{metriceqtEinstein})
\& (\ref{scalareqtEinstein})). The differences will be crucial in reconciling the
Einstein frame and the Jordan frame viewpoints on the problem of motion.

Linearizing the metric perturbations, but not the scalar field perturbation,
we obtain the following metric equations:
\begin{eqnarray}
\label{metriceqtJordan2}
&& \nabla^2 \Psi = {\Omega^{-2}} \Big( 4\pi G  \rho 
- \sfrac14 h \nabla_0 \varphi \nabla^0 \varphi
+ \sfrac14 h \nabla_i \varphi \nabla^i \varphi  \nonumber \\
&& \quad \quad \quad   +\sfrac12\Omega^4 V
+ \sfrac12  \nabla^2 \Omega^2  \Big)
\nonumber \\
&& \partial_0 \partial_i \Psi = -\sfrac12 \Omega^{-2} h \nabla^0\varphi
\nabla_i \varphi - \sfrac12 \Omega^{-2} \nabla_i \nabla^0 \Omega^2
\nonumber \\
&& (\partial_i \partial_j - \sfrac13 \delta_{ij} \nabla^2) (\Psi - \Phi)
= \nonumber \\ 
&& \quad \quad \quad \Omega^{-2} \left( h \nabla_i \nabla^j \varphi - 
\sfrac13 {\delta_i}^j h \nabla_k \varphi \nabla^k \varphi \right)
\nonumber \\ 
&& \quad \quad \quad
+ \Omega^{-2} \left( \nabla_i \nabla^j \Omega^2 - \sfrac13 {\delta_i}^j
\nabla^2 \Omega^2 \right)
\nonumber \\
&& 4 \nabla^2 \Psi - 2 \nabla^2 \Phi - 6 \partial_0^2 \Psi = \nonumber \\
&&  \quad \quad \quad \Omega^{-2} \left( 8\pi G \rho + h \nabla_\alpha \varphi \nabla^\alpha \varphi+ 4 \Omega^4 V \right) \nonumber \\
&&  \quad \quad \quad  + 3 \Omega^{-2} \Box \Omega^2
\end{eqnarray}
where the last equation comes from the space-time trace.
Combining this trace equation with the scalar field equation (Eq. (\ref{scalareqtJordan})),
we obtain:
\begin{eqnarray}
\label{scalareqtJordan2}
\Omega^{-2} \Box \varphi + \Omega^{-4} 
{\partial \Omega^2 \over \partial \varphi} \partial_\mu \varphi \partial^\mu \varphi
= {\partial V \over \partial \varphi}
-  \Omega^{-4} {\partial {\,\rm ln\,} \Omega^2 \over \partial \varphi} 4 \pi G\rho \; .
\end{eqnarray}
This is nothing other than the conformally transformed version
of the scalar field equation in Einstein frame (
Eq. (\ref{scalareqtEinstein})). Ignoring time derivatives, metric perturbations
and using Eqs. (\ref{alphaphi}) \& (\ref{alphastar}), we obtain
\begin{eqnarray}
\label{scalarJordan}
\nabla^2 \varphi = {\partial V \over \partial \varphi}
+ \alpha \, 8 \pi G \, \rho \; .
\end{eqnarray}
This scalar field equation has exactly the same form as
its Einstein frame counterpart Eq. (\ref{scalarEinstein}), and therefore
the chameleon mechanism works in the same way.
Let us stress that the important chameleon nonlinearity is retained in Eq. (\ref{scalarJordan}),
namely that the potential $V$ is not linearized. For instance, it could have the form
of $1/\varphi^n$ where $n$ is positive (see Fig. \ref{potential}).

The metric equations Eq. (\ref{metriceqtJordan2}) can be simplified by
invoking the same set of assumptions we used in the Einstein frame i.e.
time derivatives are small,
$\varphi \ll 1$, and
$V$ can be ignored both inside the object of interest (because
$G \rho \gg V$) and outside at the radius of interest $r$:
\begin{eqnarray}
\label{metriceqtJordan3}
&&\nabla^2 (\Psi + \Phi)/2 = 4\pi G \Omega^{-2} \rho \simeq 4\pi G \rho \nonumber \\
&&\nabla^2 (\Psi - \Phi) = \Omega^{-2} \nabla^2 \Omega^2 \simeq - 2 \nabla^2 q \; .
\end{eqnarray}
This is the analog of Eq. (\ref{Einsteinframe1orderB}) in Einstein frame.

Employing the same decomposition of fields into
contributions from environment (subscript $0$) and object (subscript $1$) as before,
\begin{eqnarray}
\label{decomposeJordan}
\Phi = \Phi_0 +  \Phi_1 (r) \, , \,
\Psi = \Psi_0 +  \Psi_1 (r) \, , \,
\varphi = \varphi_0 + \varphi_1 (r)
\end{eqnarray}
we obtain the following solutions to Eqs. (\ref{scalarJordan}) \& 
(\ref{metriceqtJordan3}):
\begin{eqnarray}
\label{fieldsolutionsJordan}
\Phi_0 = \Phi_0 ({0}) + \di_i \Phi_0 ({0}) x^i
\quad , \quad \Phi_1 = -{GM \over r} + q_1 \nonumber \\
\Psi_0 = \Psi_0 ({0}) + \di_i \Psi_0 ({0}) x^i
\quad , \quad \Psi_1 = -{GM \over r} - q_1 \nonumber \\
\varphi_0  = \varphi_* + \di_i \varphi_0(0) x^i
\quad , \quad
\varphi_1  = - \epsilon \, \alpha {2GM \over r}
\end{eqnarray}
where $q_1$ is defined below (Eq. (\ref{qdecompose})).
These solutions are obtained under the same set of assumptions enumerated
at the end of \S \ref{einstein}, namely we can find a radius $r$ outside the object
which is:
1.~small enough such that
second derivatives of the environmental fields can be ignored;
2.~large enough such that the monopole contribution from the object itself dominates.
The solution for $\varphi$ is non-trivial: see discussion in
\S \ref{einstein} on the chameleon mechanism. The
quantity $\epsilon$ signifies whether chameleon screening occurs:
it equals unity if there is no screening, and becomes small if
there is screening (see Eqs. (\ref{thinshell}) \& (\ref{nothinshell})).

Note that we could have easily obtained these expressions
by applying the small conformal transformation Eq. (\ref{smallconformal})
to the corresponding Einstein frame results.

The next step is to use these solutions to compute the gravitational force
by way of a surface integral at radius $r$ (Eq. (\ref{force})).   
Here, we need to be careful in
defining the relevant pseudo energy-momentum tensor.
The Jordan frame metric equation (Eq. (\ref{metriceqtJordan}))
can be rewritten in the form of Eq. (\ref{G1eqt}) with 
\begin{eqnarray}
\label{tmunuJordan}
&& t_\mu {}^\nu = \Omega^{-2} ({T^m}_\mu {}^\nu +
{{T^\varphi}_\mu}^\nu) - {1\over 8\pi G} {G^{(2)}}_\mu {}^\nu 
\nonumber \\
&& \quad \quad + {\Omega^{-2} \over 8\pi G}
\left[\nabla_\mu \nabla^\nu \Omega^2 
- {\delta_\mu}^\nu \Box \Omega^2 \right] \; .
\end{eqnarray}

As usual, we ignore the exterior matter energy-momentum at radius $r$.
We have already computed the relevant surface integral
involving $G^{(2)}$ (Eq. (\ref{G2integrate})).
We need to redo the surface integral over the scalar field
energy-momentum because unlike in Einstein frame, our scalar field here
is non-canonical. We also need to consider a scalar field
contribution that is unique to Jordan frame (second line of Eq. (\ref{tmunuJordan})).

Eqs. (\ref{TmunuscalarJordan}) together with (\ref{hdef}) tell us that
\begin{eqnarray}
&& \Omega^{-2} {{T^\varphi}_i}^j = \frac{1}{8\pi G} \Big\{ \di_i \varphi \, \di^j \varphi
- {\delta_i}^j \left[ \sfrac12 \di_k \varphi \, \di ^k  \varphi + V \right]
\nonumber \\
&& \quad \quad - 6 \left[ \di_i q \, \di ^j q - {\delta_i}^j \sfrac12
\di_k q \, \di ^k q \right] \Big\}
\end{eqnarray}
where we have ignored time derivatives, and used
$\Omega^2 \equiv 1 - 2q$ with $|q| \ll 1$. 
As expressed in Eq. (\ref{alphastar}), we can approximate $\partial q/\partial\varphi$
as constant, which means we can Taylor expand $q(\varphi)$ to first order:
\begin{eqnarray}
\label{qdecompose}
&& q = q_0 + q_1 \quad \quad {\,\rm where} \nonumber \\
&& q_0 \equiv q(\varphi_*) + \alpha
(\varphi_0 - \varphi_*) \, , \quad
q_1 \equiv \alpha \,  \varphi_1
\end{eqnarray}
where $\varphi_*$ is some typical scalar field value exterior to the object,
and $\varphi_0$ and $\varphi_1$ are as defined in Eq. (\ref{fieldsolutionsJordan}).

Applying the same arguments that go into obtaining the
Einstein frame result Eq. (\ref{Tscalarintegrate}), we find
\be
\label{scalarJordanintegrate}
- \oint dS_j \Omega^{-2} {{T^\varphi}_i}^j = \frac{r^2}{2G} \Big[ -  \partial_i \varphi_0 
{\partial \varphi_1 \over \partial r} + 6 \, \partial_i q_0 {\partial q_1 \over
\partial r} \Big] \; .
\ee

Lastly, the contribution from the
uniquely Jordan frame terms
(last line of Eq. (\ref{tmunuJordan})) works out to be
\begin{eqnarray}
\label{uniqueJordanintegrate}
- \oint dS_j {\Omega^{-2} \over 8\pi G}
\left[ \nabla_i \nabla^j \Omega^2 - {\delta_i}^j
\Box \Omega^2 \right]
= {r^2 \over 3G} \times \nonumber \qquad \quad && \\
\left[ -4 \partial_i q_0 {\partial q_1 \over \partial r}
+ 4 \partial_i q_0 {\partial \Psi_1 \over \partial r}
- \partial_i q_0 {\partial \Phi_1 \over \partial r}
- \partial_i \Phi_0 {\partial q_1 \over \partial r} \right] .&&
\end{eqnarray}

Combining together all these contributions
from Eqs. (\ref{G2integrate}), (\ref{scalarJordanintegrate})
\& (\ref{uniqueJordanintegrate}), we find
\begin{eqnarray}
\dot P_i   =   - \oint dS_j t_i {}^j  \nonumber \qquad \qquad \qquad \qquad \qquad \qquad \qquad &&   \\ 
= {r^2 \over 2G} \Big[ - \partial_i \Phi_0 {\partial \over \partial r} ( \sfrac43 \Psi_1 + \sfrac23 \Phi_1
+ \sfrac23 q_1 ) - \partial_i \varphi_0 {\partial \varphi_1 \over \partial r} \nonumber && \\
+ \partial_i q_0 {\partial \over \partial r} ( \sfrac{10}{3} q_1 + \sfrac83 \Psi_1 - \sfrac23 \Phi_1)
\Big] \nonumber \qquad \qquad && \\
= {r^2 \over 2G} \Big[  
- \partial_i (\Phi_0 - q_0) {\partial (\Phi_1 + \Psi_1) 
\over \partial r}
- \partial_i \varphi_0 {\partial\varphi_1 \over \partial r} \Big]  , \quad &&
\end{eqnarray}
where for the last equality we have used the fact that $\Phi_1 - \Psi_1 = 2 q_1$ 
from Eq. (\ref{fieldsolutionsJordan}).

Substituting the solutions for $\Phi_1$, $\Psi_1$ and $\varphi_1$ 
from Eq. (\ref{fieldsolutionsJordan}), and equating
$\dot P_i$ with $M \ddot X^i$ where $X^i$ is the center of mass
coordinate (Eq. (\ref{Pdotz})), we obtain
\begin{eqnarray}
\label{Jordanmain}
M \ddot X^i = - M \partial_i (\Phi_0 - q_0) - \epsilon
\alpha  M \partial_i \varphi_0 \; .
\end{eqnarray}
This result is consistent with the Einstein frame equation
of motion Eq. (\ref{Einsteinmain}) once we recognize that
$\tilde \Phi_0 = \Phi_0 - q_0$ (see Eq. (\ref{smallconformal})).
In fact, we can simplify this further by using 
Eq. (\ref{qdecompose}):
\begin{eqnarray}
\label{Jordanmain2}
M \ddot X^i = - M \, \partial_i \Phi_0 + (1 - \epsilon)
\alpha  M \, \partial_i \varphi_0
\end{eqnarray}
which reproduces Eq. (\ref{Jordanmain2main}).

We can simplify this further 
if we make an additional assumption about how the environmental
fields $\Phi_0$ and $\varphi_0$ are sourced, which does not necessarily hold
in general. That both $\Phi_0 - q_0$ and $\varphi_0$ are sourced primarily
by the density:
\begin{eqnarray}
\label{sourceenvironmentJordan}
\nabla^2 (\Phi_0 - q_0) = 4\pi G \rho \quad , \quad
\nabla^2 \varphi_0 = \alpha \, 8\pi G \rho
\end{eqnarray}
where $\rho$ is the environmental density. These are the analogs of
Eq. (\ref{sourceenvironment}) in the Einstein frame. With this assumption,
we can see that
\begin{eqnarray}
\partial_i \varphi_0 \left[ 1 + 2 \alpha^2 \right] = 2 \alpha \, 
 \partial_i \Phi_0
\end{eqnarray}
which then implies Eqs. (\ref{Jordanmain3main}) and (\ref{twomasses2}).


\section{Total force in the Vainshtein case}
\label{forceVainshtein}

Consider the effective 4D description of DGP. It is a scalar-tensor theory, where the scalar Lagrangian in Einstein frame is given by Eq.~(\ref{DGP}).
Assuming that we have a spherical object, we want to compute the total force acting on it as
\be
\dot P_i =  - \oint_S t_i {}^j  dS_j \; ,
\ee
where $S$ is a sphere surrounding the object, possibly deep in the non-linear regime for $\pi$.
The effective stress-tensor $t_i {}^j$ receives three kinds of contributions:
\begin{enumerate}
\item A gravitational one, bilinear in $\Phi$ and $\Psi$, proportional to ${G^{(2)}}_i {}^j$;
\item A free scalar-like one, schematically of the form $\di \pi \, \di \pi$, coming from the quadratic part of eq.~(\ref{DGP});
\item A cubic one, of the form $\di \pi \, \di \pi \, \di^2 \pi$, coming from the cubic part of eq.~(\ref{DGP}).
\end{enumerate}
The first two contributions are already dealt with in \S \ref{GR}, \S \ref{einstein}, and we need not compute them again here. We will focus instead on the third contribution, which is  the intrinsically new feature introduced by the Vainshtein effect. 

The cubic part of the stress-energy tensor is
\bea \label{T3}
T^{(3)}_{\mu\nu} & = & \frac{2\mpl^2}{m^2}\Big[ 2 \, \di_\mu \pi \di_\nu \pi \Box \pi 
- \big( \di_\mu \pi \di_\nu(\di \pi)^2 + \di_\nu \pi \di_\mu(\di \pi)^2\big) \nonumber  \\
& + &  \eta_{\mu\nu} \di_\alpha \pi \di ^\alpha (\di \pi)^2 \Big] \; .
\eea
This comes from varying the cubic part of the $\pi$ action (\ref{DGP}) w.r.t.~to the metric. (The variation of $\Box \pi$ is straightforward upon rewriting $\Box \pi = \frac{1}{\sqrt{-g}} \di_\mu (\sqrt{-g} g^{\mu\nu} \di_\nu \pi)$.)
We want to compute
\be
\oint_S {T^{(3)}}_i {}^j  dS_j \; .
\ee
We split the total $\pi$ field as the sum of an environmental one and of that created by the object
\be
\pi = \pi_0 + \pi_1 \; .
\ee
As discussed in \S \ref{strongcoupling}, this splitting is well defined even in the non-linear regime, as long as the external field $\pi_0$ can be approximated as a constant gradient field:
\be \label{pi0}
\vec \nabla \pi_0 \simeq {\rm const}.
\ee
Also, assuming spherical symmetry of the source, we restrict to a radial $\pi_1$:
\be
\pi_1 = \pi_1(r) \; .
\ee
Given (\ref{pi0}), second derivatives in (\ref{T3}) have to act on $\pi_1$. Also, upon integrating over the sphere the only surviving terms in (\ref{T3}) are those involving one $\pi_0$ field and two $\pi_1$'s---all the others integrate to zero because of spherical symmetry. 
We thus have
\bea
\oint_S {T^{(3)}}_i {}^j  dS_j & = & \frac{2\mpl^2}{m^2}  \oint_S \Big[ 2 \,  \di_i \pi_0 \big( \di_j \pi_1 \, \nabla^2 \pi_1 -  \di_k \pi_1 \, \di_j\di_k \pi_1 \big) \nonumber \\
& - & 2 \,  \di_k \pi_0 \big( \di_i \pi_1 \,  \di_j \di_k \pi_1 - \delta_{ij}  \, \di_l \pi_1 \, \di_k \di_l \pi_1 \big)  \nonumber \\
&+ & (i \leftrightarrow j) \, \Big] dS_j  \; .
\eea
Using
\bea
\di_i \pi_1 & = & \hat r_i \,  \pi_1' \\ 
\di_i \di_j \pi_1 & = & \big(\delta_{ij}-\hat r_i \hat r_j \big)  \frac{\pi_1'}{r} + \hat r_i  \hat r_j \, \pi_1'' 
\eea
(primes denote derivatives w.r.t.~$r$) we finally get
\bea
\oint_S {T^{(3)}}_i {}^j  dS_j & = &\frac{2\mpl^2}{m^2} \oint_S dS \, \frac1r \Big[ \di_i \pi_0 \, 2 {\pi_1'} ^2 \nonumber \\
& & + \, \di_j \pi_0 \, 6 {\pi_1'} ^2\, \hat r_i \hat r_j \Big]   \nonumber \\
& = & \frac{8\pi \mpl^2}{m^2} \, 4 R \,  {\pi_1'} ^2 \,  \di_i \pi_0 \; ,
\eea
where $R$ is the radius of $S$. To this, we have to add the GR contribution (\ref{MPhi0}), as well as the 
scalar contribution coming from the quadratic part of the Lagrangian---eq.~(\ref{Tscalarintegrate}), corrected by a factor $6$ to make up for the non-canonical normalization of $\pi$:
\be \label{DGPforce}
\dot {\vec P} = - M \vec \nabla \tilde \Phi_0 - 8 \pi \mpl^2 \, R^2 \Big[ 3 \pi_1' + \frac{4}{m^2} \frac{{\pi_1'}^2}{R} \Big] \,  \vec \nabla \pi_0 \; .
\ee

We now have to relate $\pi'_1$ to the mass of the source and to $R$. As discussed in \S \ref{strongcoupling}, the field equation deriving from (\ref{DGP}) can be rewritten as a Gauss law,
\be
\oint \vec J \cdot d \vec S = M \; ,
\ee
where for a spherically symmetric $\pi_1(r)$ the current $\vec J$ is \cite{NR}
\be
\vec J = \mpl^2  \Big[ 6 \pi_1 ' + \frac{8}{m^2} \, \frac{{\pi_1'} ^2 }{r}\Big] \hat r \; .
\ee
This is precisely the combination appearing in eq.~(\ref{DGPforce}).
We thus see that the total force acting on the object is simply
\be
\dot {\vec P} = - M \vec \nabla \tilde \Phi_0 - M \vec \nabla \pi_0 \; ,
\ee
independently of $R$, that is, independently of whether  the sphere $S$ is in the linear or non-linear region. This equation is in Einstein frame. Transforming to Jordan frame gives simply:
\be
\dot {\vec P} = - M \vec \nabla \Phi_0
\ee


\section{Results in an FRW Universe}
\label{app:FRW}

Here, we wish to provide extensions of our
results to an expanding universe. We will do this in Einstein frame.
Jordan frame results can be obtained by making a conformal transformation
(Eqs. (\ref{conformaltransf}), (\ref{alphaphi}), (\ref{smallconformal})).

The metric is
\be
\label{FRW}
ds^2 = a(\eta)^2 \left[ - (1 + 2 \tilde\Phi) d\eta^2 + (1 - 2 \tilde\Psi) \delta_{ij} dx^i dx^j \right] .
\ee
We will use ${\,}'$ for derivative with respect to conformal time $\eta$.

The components of the Einstein tensor up to first order in metric perturbations are
\begin{eqnarray}
&& {\tG_0}^0 = - {3\over a^2} \left({a' \over a}\right)^2 - {2\over a^2} \nabla^2 \tilde\Psi 
+ {6\over a^2} {a' \over a} \left( {a' \over a} \tilde \Phi + \tPsi' \right) \nonumber \\
&& {\tG_i}^0 = - {2 \over a^2} \left( {a' \over a} \partial_i \tPhi + \partial_i \tPsi' \right) \nonumber \\
&& {\tG_0}^i = {2 \over a^2} \left( {a' \over a} \partial_i \tPhi + \partial_i \tPsi' \right) \nonumber \\
&& {\tG_i}^j = \delta_{ij} \Big[ - {1\over a^2} \left( 2 {a'' \over a} - {{a'}^2 \over a^2} \right)
+ {2\over a^2} \left( 2 {a'' \over a} - {{a'}^2 \over a^2}\right) \tPhi \nonumber \\
&& \quad \quad - {1\over a^2} \nabla^2 (\tPsi - \tPhi) 
+ {2\over a^2} \left( \tPsi'' + {a' \over a} \left( 2 \tPsi' + \tPhi' \right) \right) \Big] \nonumber \\
&& \quad \quad + {1\over a^2} \partial_i \partial_j (\tPsi - \tPhi) \, .
\end{eqnarray}

The affine connection components are, up to first order:
\begin{eqnarray}
&& {\Gamma^0}_{00} = {a' \over a} + \tPhi' \, , \qquad{\Gamma^0}_{0i} = \partial_i \tPhi \nonumber \\
&& {\Gamma^0}_{ij} = \delta_{ij} \left[ {a'\over a} (1 - 2\tPsi - 2\tPhi) - \tPsi' \right]\nonumber \\
&& {\Gamma^i}_{00} = \partial_i \tPhi \, , \qquad
{\Gamma^i}_{0j} =  \delta_{ij} \left[ {a'\over a} - \tPsi' \right]\nonumber \\
&& {\Gamma^i}_{jk} = - \delta_{ij} \partial_k \tPsi - \delta_{ik} \partial_j \tPsi
+ \delta_{jk} \partial_i \tPsi \, .
\end{eqnarray}

Using a $\bar {\,}$ to denote the spatially homogeneous components,
we have the following zeroth-order Einstein equations
\begin{eqnarray}
&& 3 {{a'}^2 \over a^4} = 8\pi G 
\left[ {\bar{\tilde\rho}} + {1\over 2} \mpl^2 {\dot{\bar{\varphi}}}^2 + \mpl^2 V(\bar\varphi) \right]
\nonumber \\
&& - {a'' \over a^3} -  {1\over 2} {{a'}^2 \over a^4}
= 4\pi G \left[ {1\over 2} \mpl^2 {\dot{\bar{\varphi}}}^2 - \mpl^2 V(\bar\varphi) \right] .
\end{eqnarray}
The somewhat unusual appearance of factors of $\mpl^2$ arises from our choice of
action (Eq. (\ref{EF})) which makes $\varphi$ dimensionless.
Working within the same approximation as before, we
can ignore the kinetic term for $\varphi$. The potential $V$ 
consists of a constant piece (cosmological constant) plus
a $\bar\varphi$ dependent piece. The latter is constrained by the last expression of
Eq. (\ref{Einsteinframe1storder}), which tells us if $\bar\varphi$ is sitting
at equilibrium, then ${\delta V} \sim \alpha {\bar{\tilde\rho}} \bar \varphi /\mpl^2
\ll {\bar{\tilde\rho}}/\mpl^2$. This means the non-constant part of $V$ can actually
be ignored i.e.~as far as the background expansion is concerned, a theory
with a small scalar field value such as $f(R)$ should be very close to simply
having a cosmological constant \cite{Brax2008}. A possible exception is if the potential
violates $\partial V/\partial\varphi \sim \delta V/ \varphi$.

In the 
non-relativistic (no time derivatives, small peculiar velocities) plus
sub-Hubble limit, the first order Einstein equations
reduce to the analog of Eq. (\ref{Einsteinframe1orderB}):
\begin{eqnarray}
\nabla^2 \tPsi = 4\pi G a^2 \tilde\rho \quad , \quad
\nabla^2 (\tPhi - \tPsi) = 0 \, .
\end{eqnarray}
Writing $\varphi = \bar\varphi + \delta\varphi$, 
the perturbed (though not necessarily first order) scalar field equation in the 
quasi-static limit is
\begin{eqnarray}
{1\over a^2} \nabla^2 \delta \varphi = {\partial V \over \partial \delta \varphi}
+ \alpha 8\pi G \delta \tilde \rho \, .
\end{eqnarray}
analogous to Eq. (\ref{scalarEinstein}).

To implement the surface integral approach to the problem of motion,
we need some generalized concept of conservation in comoving space.
Let us make the following split: ${\tG_\mu}^\nu = {{\bar \tG}_\mu} {}^\nu
+ \delta {\tG_\mu}^\nu$, where ${{\bar \tG}_\mu} {}^\nu$ is the FRW background piece.
The perturbed (not necessarily small) piece $\delta {\tG_\mu}^\nu$
can be written as the sum of the first order part $\delta {{\tG^{(1)}}_\mu}^\nu$
plus the rest. It can be shown that in the quasi-static, sub-Hubble limit,
we have the following identities:
\begin{eqnarray}
&& \partial_\nu (a^3 \delta {{\tG^{(1)}}_0}^\nu) = 0 \nonumber \\
&& \partial_\nu (a^4 \delta {{\tG^{(1)}}_i}^\nu) = a^4 \partial_i \tPhi
\left[ {{\bar \tG}_0}^{\,0} - {1\over 3} {{\bar \tG}_k}^{\,k} \right] \, ,
\end{eqnarray}
where the first expression assumes the quasi-static/non-relativistic, sub-Hubble limit
while the second expression is exact. 

The perturbed Einstein equation, keeping all the nonlinearities but
subtracting out the background FRW piece, can be written as
\begin{eqnarray}
&& \delta {{\tG^{(1)}}_\mu}^\nu = 8 \pi G {\delta \tt_\mu}^\nu \nonumber \\
&& \delta {\tt_\mu}^\nu \equiv \delta {\tT_\mu}^\nu - {1\over 8\pi G} 
(\delta {{\tG}_\mu}^\nu - \delta {{\tG^{(1)}}_\mu}^\nu ) \, .
\end{eqnarray}

We therefore have the following modified `conservation' laws of
the pseudo-energy-momentum:
\begin{eqnarray}
&& \partial_\nu (a^3 \delta {\tt_0}^\nu) = 0 \nonumber \\
&& \partial_\nu (a^4 \delta {\tt_i}^\nu) = a^4 \partial_i \tPhi \,
\left[ {{\bar \tT}_0}^{\,0} - {1\over 3} {{\bar \tT}_k}^{\,k} \right] \, .
\end{eqnarray}
Note that indexes on $\delta {\tt_\mu}^\nu$ are raised and lowered by
the background FRW metric.

The mass of our object of interest is
\begin{eqnarray}
M = - \int d^3 x a^3 \delta {\tt_0}^0 \, ,
\end{eqnarray}
where we assume the mass is dominated by the overdensity; the background
contributes very little over our volume. Note that $x$ is comoving.
The first conservation law can be used to show that $M$ is nearly
constant---as long as there is negligible flux through the surface
bounding our volume.
Define then the center of mass coordinate
\begin{eqnarray}
X^i \equiv - \int d^3 x a^3 \delta {\tt_0}^0 x^i / M \, ,
\end{eqnarray}
and the same conservation law tells us
\begin{eqnarray}
{X^i}' = - \int d^3 x a^3 \delta {\tt_0}^i / M
= \int d^3 x a^3 \delta {\tt_i}^0 / M \, ,
\end{eqnarray}
where the ${\,}'$ denotes derivative with respect to conformal time.
Finally, we can employ the second 'conservation' law
to show
\be
(a X^i {}')' = {a^4\over M} \int d^3 x \left[ - \partial_j \delta {\tt_i}^j + 
\partial_i \tPhi \left( {{\bar \tT}_0}^{\,0} - {{{\bar \tT}_k}^{\,k} \over 3} \right) \right] \, .
\ee
The first term on the right can be rewritten as a surface integral
using Gauss law. The second term, upon splitting $\tPhi = \tPhi_0 + \tPhi_1$
as usual, gives us something like
$a \partial_i \tPhi_0$ times the mass from the background, which is negligible
compared with the mass of the object.

Carrying out the surface integral, we obtain the analog of
Eq. (\ref{Einsteinmain2}):
\begin{eqnarray}
M \left[ {X^i}'' + {a' \over a} {X^i}' \right]
= - M [1 + 2\epsilon \alpha^2] \partial_i \tPhi_0 \, ,
\end{eqnarray}
which holds in the quasi-static, sub-Hubble limit.
The only difference from our earlier result is the appearance of
a Hubble drag term, as expected.

Transforming to Jordan frame, the equation is
\begin{eqnarray}
M \left[ {X^i}'' + {a' \over a} {X^i}' \right]
= - M \partial_i \Phi_0 + (1 - \epsilon) \alpha M \partial_i \varphi_0 \, ,
\end{eqnarray}
which is analogous to Eq. (\ref{Jordanmain2}).

\newcommand\spr[3]{{\it Physics Reports} {\bf #1}, #2 (#3)}
\newcommand\sapj[3]{ {\it Astrophys. J.} {\bf #1}, #2 (#3) }
\newcommand\sapjs[3]{ {\it Astrophys. J. Suppl.} {\bf #1}, #2 (#3) }
\newcommand\sprd[3]{ {\it Phys. Rev. D} {\bf #1}, #2 (#3) }
\newcommand\sprl[3]{ {\it Phys. Rev. Letters} {\bf #1}, #2 (#3) }
\newcommand\np[3]{ {\it Nucl.~Phys. B} {\bf #1}, #2 (#3) }
\newcommand\smnras[3]{{\it Monthly Notices of Royal
        Astronomical Society} {\bf #1}, #2 (#3)}
\newcommand\splb[3]{{\it Physics Letters} {\bf B#1}, #2 (#3)}

\newcommand\AaA{Astron. \& Astrophys.~}
\newcommand\apjs{Astrophys. J. Suppl.}
\newcommand\aj{Astron. J.}
\newcommand\mnras{Mon. Not. R. Astron. Soc.~}
\newcommand\apjl{Astrophys. J. Lett.~}
\newcommand\etal{{~et al.~}}

\end{document}